\documentclass[12pt]{JHEP3}
\usepackage{epsfig}

\title{ The finite-temperature chiral transition in QCD
with adjoint fermions } 
\author{Francesco Basile\\ Dipartimento di
Fisica dell'Universit\`a di Pisa, \\ Via Buonarroti 2, 56127 Pisa,
Italy.
\email{basile@sns.it} } 
\author{Andrea Pelissetto\\
Dipartimento di Fisica dell'Universit\`a di Roma ``La Sapienza'' and
INFN, \\ Piazzale Moro 2, 00185 Roma, Italy.
\email{Andrea.Pelissetto@roma1.infn.it} } 
\author{Ettore Vicari \\
Dipartimento di Fisica dell'Universit\`a di Pisa and INFN, \\ Via
Buonarroti 2, 56127 Pisa, Italy.
\email{vicari@df.unipi.it} }

\abstract{ We study the nature of the finite-temperature chiral
transition in QCD with $N_f$ light quarks in the adjoint representation (aQCD).
Renormalization-group arguments
show that the transition can be continuous if a stable fixed point
exists in the renormalization-group flow of the corresponding
three-dimensional $\Phi^4$ theory with a complex $2N_f\times 2N_f$
symmetric matrix field and symmetry-breaking pattern ${\rm
SU}(2N_f)\rightarrow {\rm SO}(2N_f)$.  This issue is investigated by
exploiting two three-dimensional perturbative approaches, the massless
minimal-subtraction scheme without $\epsilon$ expansion and a massive
scheme in which correlation functions are renormalized at zero
momentum. We compute the renormalization-group functions in the two
schemes to five and six loops respectively, and determine their
large-order behavior.

The analyses of the series show the presence of a stable
three-dimensional fixed point characterized by the symmetry-breaking
pattern ${\rm SU}(4)\rightarrow {\rm SO}(4)$. This fixed point does
not appear in an $\epsilon$-expansion analysis and therefore does not
exist close to four dimensions.  The finite-temperature chiral
transition in two-flavor aQCD can therefore be continuous; in this
case its critical behavior is determined by this new SU(4)/SO(4)
universality class.  One-flavor aQCD may show a more complex phase
diagram with two phase transitions.  One of them, if continuous,
should belong to the O(3) vector universality class.  }


\begin{document}

\section{Introduction}
\label{intro}

The thermodynamics of Quantum Chromodynamics (QCD) is characterized by
a transition from a low-temperature hadronic phase, in which chiral
symmetry is broken, to a high-temperature phase with deconfined quarks
and gluons (quark-gluon plasma), in which chiral symmetry is
restored. See, e.g.,
Refs~\cite{Wilczek-00,Karsch-02,Rajagopal-95,KL-03} for recent
reviews.  Although deconfinement and chiral symmetry restoration are
apparently related to different nonperturbative mechanisms, they seem
to be somehow coupled in QCD. Indeed, lattice computations show that
the Polyakov loop has a sharp increase around the critical temperature
where the chiral condensate vanishes \cite{Karsch-02}.  However, the
interplay between the two effects is not clear yet.

Insight into this question may be gained by investigating QCD-related
models, such as ${\rm SU}(N_c)$ gauge theories with $N_f$ Dirac
fermions in the adjoint representation (aQCD). In aQCD with $N_f$
massless flavors the chiral symmetry group is ${\rm SU}(2N_f)$, which
is expected to be spontaneusly broken to ${\rm SO}(2N_f)$ at low
temperature \cite{Peskin-80,VW-84,LS-92,SV-95,KSTVZ-00}, due to the
presence of a nonzero quark condensate.  In addition, unlike QCD with
fermions in the fundamental representation, aQCD is also invariant
under global ${\mathbb Z}_{N_c}$ transformations corresponding to the
center of the SU($N_c$) gauge group, as is the case in pure ${\rm
SU}(N_c)$ gauge theories. This symmetry breaks down in the
high-temperature deconfined phase.  Therefore, in aQCD one expects two
finite-temperature transitions: a deconfinement transition at $T_d$
associated with the breaking of the ${\mathbb Z}_{N_c}$ symmetry, and
a chiral transition at $T_c$ characterized by the symmetry-breaking
pattern ${\rm SU}(2N_f) \rightarrow {\rm SO}(2N_f)$. Of course, it is
also possible to have only one transition if the corresponding
critical temperatures coincide.  We should mention that this issue is
interesting only for the values of $N_f$ for which aQCD is
asymptotically free.  Since the first coefficient $b_0$ of the $\beta$
function $\beta(g)=-b_0 g^4 + O(g^6)$ of SU($N_c$) gauge theories with
$N_f$ adjoint Dirac fermions is $b_0=(11- 4N_f)N_c/(24\pi^2)$, aQCD is
asymptotically free only for $N_f<11/4$, i.e., for $N_f = 1$ and $N_f
= 2$.

Monte Carlo simulations of ${\rm SU}(2)$ \cite{KPSW-85,Kogut-87} and
SU(3) \cite{KL-99} gauge theories with two adjoint fermions show that
the deconfinement and chiral transitions are well separated with
$T_d<T_c$.  In the case of three-color aQCD, the Monte Carlo
simulations reported in Ref.~\cite{KL-99} show actually that the ratio
$T_c/T_d$ is rather large, $T_c/T_d\simeq 8$, suggesting a rather weak
interplay between the corresponding underlying mechanisms.  They
provide a rather clear evidence that the deconfinement transition
associated with the center symmetry ${\mathbb Z}_3$ is of first order.
Moreover, the available data at the chiral transition are apparently
consistent with a continuous transition. However, they should be
considered as rather preliminary and not conclusive, since a careful
analysis of finite-size effects and of the approach to the continuum
limit has not been done yet.  The phase diagram of two-color aQCD in
the temperature--chemical-potential plane has been recently discussed
in Ref.~\cite{ST-04}.

In this paper we investigate the nature of the finite-temperature
chiral transition in aQCD, i.e. in four-dimensional ${\rm SU}(N_c)$
gauge theories with adjoint fermions, using renormalization-group (RG)
arguments.  Our study parallels the ones reported in
Refs.~\cite{PW-84,BPV-03}, in which the nature of the
finite-temperature transition in QCD with fermions in the fundamental
representation was investigated.  We consider effective
three-dimensional Landau-Ginzburg-Wilson (LGW) $\Phi^4$ theories for
the low-momentum critical modes associated with the bilinear quark condensate,
which are described by a symmetric complex $2N_f\times 2N_f$ matrix
field, and look for stable fixed points (FPs) that may be associated
with the symmetry-breaking pattern relevant for aQCD: ${\rm
SU}(2N_f)\rightarrow {\rm SO}(2N_f)$ [if the axial U(1) symmetry is
effectively restored at $T_c$ the symmetry-breaking pattern would be
${\rm U}(2N_f)\rightarrow {\rm O}(2N_f)$].  If such a stable FP does
not exist the transition is of first order; otherwise, the transition
may be continuous or of first order, depending whether the system is
or is not in the attraction domain of the stable FP.  We study the RG
flow of the LGW theories in two field-theoretical (FT) perturbative
approaches: the minimal-subtraction scheme without $\epsilon$
expansion (in the following we will indicate it as $3d$-$\overline{\rm
MS}$ scheme) \cite{SD-89} and a massive zero-momentum (MZM)
renormalization scheme \cite{Parisi-80}.  In the $3d$-$\overline{\rm
MS}$ scheme one considers the massless (critical) theory in dimensional
regularization \cite{tHV-72}, determines the RG functions from the
divergences appearing in the perturbative expansion of the correlation
functions, and finally sets $\epsilon\equiv 4-d=1$ without expanding in powers
of $\epsilon$ (this scheme therefore differs from the standard
$\epsilon$ expansion \cite{WF-72}).  In the MZM scheme one considers
instead the three-dimensional massive theory, corresponding to the
disordered (high-temperature) phase, and determines the
renormalization constants from zero-momentum correlation functions.
We compute the $\beta$ functions perturbatively to five loops in the
$3d$-$\overline{\rm MS}$ scheme and to six loops in the MZM
scheme. They are resummed by using a conformal-mapping method
\cite{LZ-77,ZJbook} taking into account their large-order behavior,
determined by means of the standard semiclassical analysis of 
instanton solutions.
Comparison of the results of these two perturbative schemes provides a
nontrivial check of the reliability of our conclusions.

We briefly summarize our main results.  In both $3d$-$\overline{\rm
MS}$ and MZM schemes the three-dimensional SU(4) LGW $\Phi^4$ theory
relevant for aQCD with two Dirac flavors shows a stable FP, that
corresponds to a new three-dimensional universality class
characterized by the symmetry-breaking pattern ${\rm SU}(4)\rightarrow
{\rm SO}(4)$.  The corresponding critical exponents are $\eta\approx
0.2$ and $\nu\approx 1.1$.  Note that this FP does not appear in an
$\epsilon$-expansion analysis ($\epsilon\equiv 4-d$) and therefore
does not exist close to four dimensions.  It is found only in
genuinely three-dimensional analyses.  This FT result implies that the
finite-temperature chiral transition in two-flavor aQCD may be
continuous. In this case, it belongs to the above-mentioned
three-dimensional SU(4)/SO(4) universality class.  However, this does
not exclude a first-order transition for systems that are outside the
attraction domain of the stable FP.  Note that, although SU(4) is
locally isomorphic to SO(6), this universality class is definitely
different from the vector O(6) one, whose symmetry-breaking pattern is
SO(6)$\rightarrow$SO(5).  No stable FP is found for the U(4) LGW
$\Phi^4$ theory which should be relevant in the case the axial U(1)
symmetry is effectively restored at $T_c$; in this case, the
transition would be of first order.  For $N_f=1$ the phase diagram may
be more complex because the phase diagram of the 
corresponding SU(2)-symmetric effective
theory has several transition lines joining at a multicritical
point.  In the parameter region relevant for aQCD, the possible phase diagrams 
have one or two phase transitions.  One of them would be associated with the
symmetry-breaking pattern ${\rm SU}(2)/{\mathbb Z}_2\cong{\rm
SO}(3)\rightarrow {\rm SO}(2)$. Therefore, if continuous, the
transition would belong to the standard O(3) (Heisenberg) universality
class. The second transition, which could be absent in aQCD, would
correspond to the symmetry-breaking pattern ${\mathbb Z}_2 \otimes
{\rm SO}(2) \to {\rm SO}(2) $, and, if continuous, would correspond to
the Ising universality class. Note that the symmetry breaking ${\rm
SU}(2)\to {\rm SO}(2) $ occurs in this case through two different
phase transitions.

The predicted critical behavior for two-flavor aQCD should be compared
with that of two-flavor QCD with quarks in the fundamental
representation. Also in this last case the finite-temperature chiral
transition may be continuous, although it would belong to a different
universality class, the vector O(4) universality class.  Monte Carlo
simulations of lattice QCD~\cite{CP-PACS-01,KLP-01,Karsch-02,KL-03}
seem to be consistent with a continuous transition in the O(4)
universality class, although they are not yet sufficiently precise to
be conclusive.

The paper is organized as follows.  In Sec.~\ref{effmod} we derive the
LGW $\Phi^4$ theory relevant for the finite-temperature chiral
transition in aQCD, using universality and RG arguments.  In
Sec.~\ref{rgun} we investigate the RG flow of the LGW theory with
U($N$) symmetry, which would be relevant for aQCD if the U(1) anomaly
were effectively suppressed at $T_c$. We report our perturbative
calculations in the $3d$-$\overline{\rm MS}$ and MZM renormalization
schemes and their analyses.  In Sec.~\ref{rgsun} we consider a more
general LGW theory in which the U($N$) symmetry is explicitly broken
to SU($N$).  We discuss the one-flavor case and present a perturbative
analysis for $N=4$, which is relevant for two-flavor aQCD.
The appendix contains an analysis of the vacuum structure
of the LGW $\phi^4$ theories relevant for aQCD.

\section{The effective Landau-Ginzburg-Wilson model at the chiral transition}
\label{effmod}

In the vanishing quark-mass limit, the fermionic part of the QCD
Lagrangian with $N_f$ adjoint Dirac fermions is given by ${\cal L}
=\sum_f \bar{\psi}_f\gamma_\mu D_\mu \psi_f$, where
$D_\mu^{ab}=\partial_\mu \delta^{ab} + i A_\mu^c f_c^{ab}$ and
$f_c^{ab}$ are the generators of the adjoint representation, i.e.  the
structure constants. Using the antisymmetry of the structure costants,
one may rewrite the Lagrangian in terms of two-component Weyl spinors
as ${\cal L} = i \sum_{i=1}^{2N_f} w_i^\dagger \sigma_\mu D_\mu w_i$,
where $\sigma_\mu\equiv ({\rm Id},\vec{\sigma})$, ${\rm Id}$ is the
identity matrix, and $\sigma_i$ are the Pauli matrices; see, e.g.,
Refs. \cite{LS-92,SV-95,KSTVZ-00}.  The actual symmetry is ${\rm
U}(2N_f)\cong {\rm U}(1)_A \otimes [{\rm SU}(2N_f)/{\mathbb
Z}_{2N_f}]$, which is larger than the symmetry ${\rm U}(N_f)_R\otimes
{\rm U}(N_f)_L$ of QCD with fermions in the fundamental
representation. The U(1)$_A$ subgroup is anomalous at the quantum
level and thus the symmetry reduces to ${\rm SU}(2N_f)$.

At $T=0$ the symmetry is expected to be spontaneously broken due to a
nonzero quark condensate $\langle \bar{\psi} \psi \rangle$.  As a
consequence of the Pauli principle, the quark bilinear condensate must
belong to the symmetric second-rank tensor representation of ${\rm
SU}(2N_f)$, which has dimension $2N_f^2+N_f$. Condensation along one
of its directions gives rise to the symmetry breaking
\begin{equation}
{\rm SU}(2N_f) \rightarrow {\rm SO}(2N_f)
\label{sym1}
\end{equation}
and to $2N_f^2+N_f-1$ Goldstone modes.  See, e.g.,
Refs.~\cite{LS-92,SV-95,KSTVZ-00} for more details.  With increasing
the temperature, a phase transition characterized by the restoring of
the chiral symmetry is expected at a given $T_c$; above $T_c$ the
quark condensate vanishes.  Therefore, the finite-temperature phase
transition is characterized by the symmetry-breaking pattern
(\ref{sym1}) and a complex symmetric $2N_f\times 2N_f$ matrix order
parameter $\Phi_{ij}$.  In the case the U(1) symmetry is effectively
restored at $T_c$, the relevant symmetry-breaking pattern would be
\begin{equation}
{\rm U}(2N_f) \rightarrow {\rm O}(2N_f).
\label{sym2}
\end{equation}
This possibility is however rather unlikely. Indeed, instanton
calculations in the high-temperature phase~\cite{GPY-81} suggest that
the axial U(1) symmetry is not restored at $T_c$, in analogy with what
happens in lattice QCD with fermions in the fundamental representation
\cite{anomalyMC}, and as also suggested by the finite-temperature
behavior of the topological
susceptibility in the pure SU($N_c$) gauge theories, see, e.g.,
Ref.~\cite{DPV-04}.

In order to investigate the nature of the finite-temperature
transition in aQCD with $N_f$ light flavors, we follow the
reasoning already applied in Refs.~\cite{PW-84,BPV-03} to the study
of the finite-temperature transition in QCD with light fermions in
the fundamental representation.

\begin{itemize}

\item[(i)] 
Let us first assume that the phase transition at $T_c$ is
continuous (second order) for vanishing quark masses.  In this case
the critical behavior should be described by an effective
three-dimensional (3-$d$) theory. Indeed, the length scale of the
critical modes diverges approaching $T_c$, becoming eventually much
larger than $1/T_c$, which is the size of the euclidean ``temporal''
dimension at $T_c$.  Therefore, the asymptotic critical behavior must
be associated with a 3-$d$ universality class characterized by a
complex symmetric $2N_f\times 2N_f$ matrix order parameter $\Phi_{ij}$
and by the symmetry-breaking pattern (\ref{sym1}) [or (\ref{sym2}) if
the U(1) symmetry is effectively restored at $T_c$].

\item[(ii)] 
According to RG theory, the existence of such universality
classes can be investigated by considering the most general LGW
$\Phi^4$ theory for a complex symmetric $N\times N$ matrix field
$\Phi_{ij}$ with $N=2N_f$ and the desired symmetry and
symmetry-breaking pattern.  The most general ${\rm U}(N)$-symmetric
LGW Lagrangian containing up to quartic terms in the potential is
\begin{eqnarray}
{\cal L}_{U(N)} = {\rm Tr}\, (\partial_\mu \Phi^\dagger) (\partial_\mu \Phi)
+r {\rm Tr}\, \Phi^\dagger \Phi 
+ {u_0\over 4} \left( {\rm Tr}\, \Phi^\dagger \Phi \right)^2
+ {v_0\over 4} {\rm Tr} \left( \Phi^\dagger \Phi \right)^2 .
\label{genun}
\end{eqnarray}
Stability requires $u_0+v_0>0$ and $Nu_0+v_0>0$.  The symmetry group
of this Lagrangian is U($N$). First, it is invariant under the
transformations $\Phi \rightarrow U\Phi U^T$, where $U$ is a unitary
${\rm U}(N)$ matrix, i.e., under the group U($N$)/${\mathbb{Z}}_2$
(the quotient ${\mathbb{Z}}_2$ is due to the fact that matrices $\pm
U$ give rise to the same transformation).  Second, it is invariant
under $\Phi \to \Phi^\dagger$.  The symmetry group is therefore
${\mathbb{Z}}_2\otimes$U($N$)/${\mathbb{Z}}_2 \cong {\rm U(}N)$. For
$v_0 > 0$, Lagrangian (\ref{genun}) shows the expected vacuum structure 
and symmetry-breaking pattern, see appendix. In the
low-temperature phase the potential is minimized by taking
\begin{equation}
\phi_{\rm min} \propto \left( 
   \begin{array}{cc}
     0 & {\rm Id} \\
     {\rm Id} & 0 
   \end{array} \right),
\end{equation}
where ${\rm Id}$ is the $N/2$-dimensional identity matrix.  Note that
$\Phi_{\rm min}$ is invariant under vector ${\rm U}(N_f)_V$
transformations, in agreement with the Vafa-Witten theorem
\cite{VW-84}.  The symmetry of the vacuum is O($N$) and thus the
symmetry breaking pattern is ${\rm U}(N)\to {\rm O(}N)$.

The axial anomaly reduces the symmetry to ${\rm SU}(N)$ 
and thus new terms
must be added. The most relevant one is proportional to $\left( {\rm
det}\,\Phi^\dagger + {\rm det}\, \Phi \right)$, which is a polynomial of
order $N$ in the field $\Phi$.  For $N>4$ such a term is
irrelevant at the transition.  Instead, for $N=2$ and $N=4$,
the determinant must be added to Lagrangian (\ref{genun}), obtaining
\begin{equation}
{\cal L}_{SU(N)} = {\cal L}_{U(N)}
+ w_0 \left( {\rm det}\, \Phi^\dagger + {\rm det}\, \Phi \right).
\label{gensun}
\end{equation}
For $N=2$ additional terms depending on the determinant of $\Phi$
should be added, in order to include all terms with at most four
fields compatible with the symmetry. They will be discussed in
Sec.~\ref{n2case}. The symmetry of Lagrangian (\ref{gensun}) is
${\mathbb Z}_2 \otimes {\rm SU}(N)$. Indeed, it is invariant under
transformations $\Phi \to U \Phi U^T$, with $U$ unitary and ${\rm
det}\, U = \pm 1$. Taking into account that $U$ and $-U$ correspond to
the same transformation, the invariance group is SU($N$). Moreover,
the model is invariant under the ${\mathbb Z}_2$ transformations $\Phi
\to \Phi^\dagger$. For $v_0 > - 3 |w_0|/2$, the vacuum structure is
identical to that discussed for the U($N$) theory, satisfying the
Vafa-Witten theorem \cite{VW-84}.  The corresponding symmetry-breaking
pattern is ${\mathbb Z}_2 \otimes {\rm SU}(N) \to {\mathbb Z}_2
\otimes {\rm SO}(N)$. Note that the symmetry group contains an
additional ${\mathbb Z}_2$ with respect to Eq.~(\ref{sym1}). This
additional invariance, related to the transformation
$\Phi\to\Phi^\dagger$, is a consequence of the hermiticity (reality)
of the effective Lagrangian and corresponds to the discrete
parity symmetry of the aQCD Lagrangian. This additional ${\mathbb
Z}_2$ is not broken at the transition, so that the relevant
symmetry-breaking pattern is indeed ${\rm SU}(N) \to {\rm SO}(N)$, as
discussed before.  It must be noted that, if a $\theta$ term is added
to the aQCD Lagrangian, this additional symmetry is not present and
the effective Lagrangian is no longer hermitian.

\item[(iii)]
The critical behavior at a continuous transition is described by the
stable FP of the theory, which determines the
universality class.  The absence of a stable FP 
indicates that the 
phase transition is not continuous.  Therefore, a necessary condition
of consistency with the initial hypothesis that the transition is
continuous, cf.~(i), is the existence of stable FPs in the theories
described by Lagrangians (\ref{genun}) and (\ref{gensun}).
If a FP exists, the transition may be either continuous, belonging to the 
universality class associated with the stable FP, or of first-order, if
the system is not in the attraction domain of the FP.

\end{itemize}

The effective theories (\ref{genun}) and (\ref{gensun}) are defined
for any $N$. However, for aQCD they are relevant only for $N$ even and
$N \le 4$, i.e., for $N = 2$ and $N = 4$.  Indeed, aQCD is
asymptotically free only for $N_f < 11/4$, i.e.,  for $N < 11/2$.

\section{Renormalization-group flow of the U($N$) LGW theory}
\label{rgun}

\subsection{Perturbative expansions}
\label{perexp}

In order to investigate the RG flow of Lagrangians (\ref{genun}) and
(\ref{gensun}), we employ two different perturbative approaches: the
MZM renormalization scheme \cite{Parisi-80} and the
$3d$-$\overline{\rm MS}$ scheme \cite{SD-89}; see
Refs.~\cite{ZJbook,PV-r} for recent reviews.  In the first case, one
considers the three-dimensional massive theory corresponding to the
disordered phase, and expresses the zero-momentum renormalization
constants in terms of zero-momentum renormalized quartic couplings. In
the second case, one considers the massless (critical) theory in
dimensional regularization within the minimal-subtraction scheme
\cite{tHV-72}.  RG functions are obtained in terms of the renormalized
couplings and of $\epsilon \equiv 4 - d$. Subsequently $\epsilon$ is
set to its physical value $\epsilon = 1$, providing a
three-dimensional scheme in which the 3-$d$ RG functions are expanded
in powers of the $\overline{\rm MS}$ renormalized quartic
couplings. This scheme differs from the standard $\epsilon$ expansion
\cite{WF-72} in which one expands the RG functions in powers of
$\epsilon$.

In order to renormalize the 
${\rm U}(N)$-invariant theory (\ref{genun}) in the $\overline{\rm
MS}$ scheme, one sets
\begin{eqnarray}
\Phi &=& [Z_\phi(u,v)]^{1/2} \Phi_R, \\
u_0 &=& A_d \mu^\epsilon Z_u(u,v) , \nonumber \\
v_0 &=& A_d \mu^\epsilon Z_v(u,v) , \nonumber
\end{eqnarray}
where $u$ and $v$ are the $\overline{\rm MS}$ renormalized quartic couplings.
The renormalization constants $Z_\phi$, $Z_u$, and $Z_v$ are
normalized so that $Z_\phi(u,v) \approx 1$,
$Z_u(u,v) \approx u$, and $Z_v(u,v) \approx v$ at tree level.  Here
$A_d$ is a $d$-dependent constant given by $A_d\equiv 2^{d-1} \pi^{d/2}
\Gamma(d/2)$.  Moreover, one defines a mass renormalization constant
$Z_t(u,v)$ by requiring $Z_t \Gamma^{(1,2)}$ to be finite when
expressed in terms of $u$ and $v$.  Here $\Gamma^{(1,2)}$ is the
one-particle irreducible two-point function with one insertion of
${\rm Tr}\, \Phi^\dagger \Phi$. 
The $\beta$ functions are computed from
\begin{equation}
\beta_u (u,v) = \mu \left. {\partial u \over \partial \mu} \right|_{u_0,v_0},
\qquad
\beta_v (u,v) = \mu \left. {\partial v \over \partial \mu} \right|_{u_0,v_0}.
\end{equation}
They have a simple dependence on $d$:
\begin{equation}
\beta_u = (d-4) u + B_u(u,v),\qquad
\beta_v = (d-4) v + B_v(u,v),
\label{Bdef}
\end{equation}
where the functions $B_u(u,v)$ and $B_v(u,v)$ are independent of $d$.
The FPs of the theory are given by the common zeroes of the $\beta$ functions.
Their stability is controlled  by the eigenvalues of the matrix
\begin{equation}
\Omega =
\left(\matrix{
{\partial \beta_u/\partial u} & {\partial \beta_u/\partial v} 
\cr
{\partial \beta_v/\partial u} & {\partial \beta_v/\partial v} 
}\right) \; .
\label{omegaun}
\end{equation}
A FP is stable if all the eigenvalues of its stability matrix have
positive real part.  
The RG functions $\eta_\phi$ and $\eta_t$
associated with the critical exponents are defined by
\begin{equation}
\eta_{\phi,t}(u,v) = \left. {\partial \ln Z_{\phi,t} \over \partial \ln \mu}
     \right|_{u_0,v_0}.
\label{etadefmsb}
\end{equation}
The functions $\eta_{\phi,t}$ are independent of $d$.
The standard critical exponents $\eta$ and $\nu$
are related to the RG functions $\eta_{\phi,t}$ 
and to the location $u^*$, $v^*$ of the FP by
\begin{equation}
\eta = \eta_\phi(u^*,v^*),\qquad
\nu = \left[ 2 + \eta_t(u^*,v^*) - \eta_\phi(u^*,v^*) \right] ^{-1}.
\label{exponents} 
\end{equation}

We computed the
$\overline{\rm MS}$ series to five loops.  For this purpose we used a
symbolic manipulation program that generates the diagrams (305
up to five loops) and computes their symmetry and group factors,
and the compilation of Feynman integrals of Ref.~\cite{KS-01}.  The
functions $B_{u,v}(u,v)$ defined in Eq.~(\ref{Bdef}) are given by
\begin{eqnarray}
B_u(u,v) &=&  
{N^2+N+8\over 4} u^2 + (N+1) u v +  {3\over 4} v^2 
-{9N^2+9N+42\over 16} u^3 
\label{bun} 
\\
&&- {11(N+1)\over 4} u^2 v 
-{5N^2+15N+92\over 32} u v^2 - {3(N+2)\over 8} v^3 
+ \sum_{i+j\geq 4} a_{ij}^u u^i v^j ,
\nonumber 
\\
B_v(u,v) &=& 3 u v +  
{2N+5\over 4} v^2 
- {5N^2+5N+82\over 16} u^2 v 
\label{bvn}
\\
&&-{11N+20\over 4} u v^2 
- {3N^2+21N+60\over 32} v^3 
+ \sum_{i+j\geq 4} a_{ij}^v u^i v^j .
\nonumber 
\end{eqnarray}
The coefficients $a_{ij}^{u}$ and $a_{ij}^{v}$ for $4\leq i+j\leq 6$,
are reported in Tables~\ref{betau} and \ref{betav} respectively.  In
order to save space, we report them numerically, although we have
their exact expressions in terms of fractions and $\zeta$ functions
with integer argument.  We do not report the series of the RG
functions $\eta_{\phi,t}$, since they will not be used in the
following.  They are available on request.

In the MZM scheme the theory is renormalized by introducing a set of
zero-momentum conditions for the one-particle irreducible two-point
and four-point correlation functions:
\begin{eqnarray}
&&\Gamma^{(2)}_{a_1a_2,b_1b_2}(p) = {\delta^2 \Gamma \over \delta
\Phi_{a_1a_2}^\dagger \delta \Phi_{b_1b_2}} =
\delta_{a_1b_1}\delta_{a_2b_2} Z_\phi^{-1} \left[
m^2+p^2+O(p^4)\right],
\label{ren1}  \\
&&\Gamma^{(4)}_{a_1a_2,b_1b_2,c_1c_2,d_1d_2}(0) =
\left. {\delta^4 \Gamma \over \delta \Phi_{a_1a_2}^\dagger 
\delta \Phi_{b_1b_2}^\dagger \delta \Phi_{c_1c_2} \delta 
\Phi_{d_1d_2}}\right|_{\rm zero \; mom.} =
\label{ren2}  \\
&&\qquad
= Z_\phi^{-2} m^{4-d} \frac{32\pi}{8+N+N^2}
\left( u U_{a_1a_2,b_1b_2,c_1c_2,d_1d_2} +  
v V_{a_1a_2,b_1b_2,c_1c_2,d_1d_2}\right),
\nonumber 
\end{eqnarray}
where $\Gamma$ is the generator of the one-particle irreducible correlation functions
(effective action), and
\begin{eqnarray}
&&U_{a_1a_2,b_1b_2,c_1c_2,d_1d_2}=
{1\over 2} \left( \delta_{a_1c_1}\delta_{b_1d_1}\delta_{a_2c_2}\delta_{b_2d_2}+
\delta_{a_1d_1}\delta_{b_1c_1}\delta_{a_2d_2}\delta_{b_2c_2} \right), 
\nonumber 
\\
&&V_{a_1a_2,b_1b_2,c_1c_2,d_1d_2}=
{1\over 2} \left( \delta_{a_1c_1}\delta_{b_1d_1}\delta_{a_2d_2}\delta_{b_2c_2}+
\delta_{a_1d_1}\delta_{b_1c_1}\delta_{a_2c_2}\delta_{b_2d_2} \right). 
\end{eqnarray}
In addition, one introduces the renormalization 
function $Z_t(u,v)$ that is defined by the 
relation
$\Gamma^{(1,2)}_{a_1a_2,b_1b_2}(0) = \delta_{a_1b_1}\delta_{a_2b_2} Z_t^{-1}$,
where $\Gamma^{(1,2)}(0)$ is the zero-momentum one-particle irreducible
two-point function with one insertion of ${\rm Tr}\, \Phi^\dagger \Phi$.
The MZM $\beta$ functions are defined by
\begin{equation}
\beta_u(u,v) = \left. m{\partial u\over \partial m}\right|_{u_0,v_0},\qquad
\beta_v(u,v) = \left. m{\partial v\over \partial m}\right|_{u_0,v_0}.
\label{betaf}
\end{equation}
The RG functions $\eta_\psi$ and $\eta_t$ associated with
the critical exponents are defined by
\begin{equation}
\eta_{\phi,t}(u,v) = \left. {\partial \ln Z_{\phi,t} \over \partial \ln m}
     \right|_{u_0,v_0}.
\label{etadefmzm}
\end{equation}

We computed the MZM RG functions to six loops.  In
this case we used the compilation of Feynman integrals of
Ref.~\cite{NMB-77} to compute the needed 1438 Feynman diagrams. 
The $\beta$ functions are given by
\begin{eqnarray}
\beta_u(u,v) &=& - u + u^2 + {4(N+1)\over N^2+N+8} u v  
+{3 \over N^2+N+8} v^2 +  \sum_{i+j\geq 3} b_{ij}^u u^i v^j ,
\label{bung} 
\\
\beta_v(u,v) &=& - v + 
{12 \over N^2+N+8} uv  + {2N+5\over N^2+N+8} v^2  
+ \sum_{i+j\geq 3} b_{ij}^v u^i v^j .
\label{bvng}
\end{eqnarray}
The coefficients $b_{ij}^{u}$ and $b_{ij}^{v}$ up to six loops,
i.e.  for $3\leq i+j\leq 7$, are reported in Tables~\ref{betaug} and
\ref{betavg} respectively.  Here we do not report the six-loop series
of the RG functions $\eta_{\phi,t}$; they are available on request.

Our calculations can be checked by considering some particular
cases.  For $N=1$ the two quartic terms are identical and one recovers
the O(2) vector model, while  for $v=0$ one obtains
the O($N^2+N$)-symmetric model. 
In these cases we can compare our perturbative expansions with those
reported in Refs.~\cite{KNSCL-91,BNGM-77,AS-95}: we find full agreement.  
In addition, for $N=2$ the model is equivalent to an
O(2)$\otimes$O(3) model \cite{Kawamura-88,Kawamura-98}. Indeed, if one sets
\begin{equation}
\Phi=\sum_{k=1}^3 (\Psi_{k1} + i \Psi_{k2}) e_k,
\label{mapo2xo3}
\end{equation}
where $(e_1,e_2,e_3)\equiv \frac{1}{2}({\rm Id},i\sigma_1,i\sigma_3)$, 
${\rm Id}$ is the $2\times 2$ identity matrix, $\sigma_i$ are
the Pauli matrices, and $\Psi_{ki}$ is a $3\times2$ real matrix, 
Lagrangian (\ref{genun}) for $N=2$ can be written as 
\begin{eqnarray}
{\cal L} &=& \int d^d x \left\{ 
  {1\over2}
\sum_{ai} \left[ \sum_\mu (\partial_\mu \Psi_{ai})^2 + r \Psi_{ai}^2 
      \right]   
\right.
\nonumber \\
&& \left.
+ {g_{1,0}\over 4!}  ( \sum_{ai} \Psi_{ai}^2)^2 
+ {g_{2,0}\over 4!}  \left[ \sum_{i,j} 
( \sum_a \Psi_{ai} \Psi_{aj})^2 - (\sum_{ai} \Psi_{ai}^2)^2 \right]
\right\} ,
\label{LGWch}
\end{eqnarray}
with 
\begin{equation}
g_{1,0} = \frac{3}{2} u_0 + \frac{3}{4}v_0,\qquad g_{2,0}=-\frac{3}{2}v_0.  
\label{mapg}
\end{equation}
This model has an explicit symmetry [O(2)$\otimes$O(3)]/${\mathbb Z}_2 
\cong {\rm U(1)}  \otimes {\rm SU(2)}/{\mathbb Z}_2 
\cong {\rm U(2)}$, as expected. It has already been
studied because it should describe transitions in frustrated
spin models with noncollinear ordering, the superfluid transition of
$^3$He, etc., see, e.g., Refs.~\cite{Kawamura-98,PV-r} and
references therein.  Six-loop series in the MZM expansion
\cite{PRV-01} and five-loop series in the $\overline{\rm MS}$
scheme~\cite{CP-04} have already been computed for generic
O(2)$\otimes$O($M$) symmetric models.  These perturbative series agree 
with ours, once we rewrite them in terms of the renormalized couplings
corresponding to the bare ones defined in Eq.~(\ref{mapg}).  
In the $\overline{\rm MS}$
scheme, the relation is 
$g_{1} = \frac{3}{2} u + \frac{3}{4}v$, $g_{2}=-\frac{3}{2}v$,
where $g_1$ and $g_2$ are the $\overline{\rm MS}$ renormalized
coupling of model (\ref{LGWch}). In the MZM scheme, the
correspondence is $\bar{g}_1 = u + v/2$ and $\bar{g}_2=-9 v/14$, where
$\bar{g}_{1,2}$ are the MZM couplings normalized as in
Ref.~\cite{PRV-01} (actually, there they were called $\bar{u}$ and
$\bar{v}$).

\subsection{Large-order behavior}

Since perturbative expansions are divergent, resummation methods must
be used to obtain meaningful results.  Given a generic quantity
$S(u,v)$ with perturbative expansion $S(u,v)= \sum_{ij} c_{ij} u^i
v^j$, we consider
\begin{equation}
S(x u,x v) = \sum_k s_k(u,v) x^k,
\label{seriesx}
\end{equation}
which must be evaluated at $x=1$. The resummation of the series
can be done by exploiting the knowledge of the large-order
behavior of the coefficients, generically given by
\begin{equation}
s_k(u,v) \sim k! \,[-A(u,v)]^{k}\,k^b\,\left[ 1 + O(k^{-1})\right].
\label{lobh}
\end{equation}
The quantity $A(u,v)$ is related to the singularity $t_s$ of the Borel
transform $B(t)$ that is nearest to the origin: $t_s=-1/A(u,v)$.  The
series is Borel summable for $x > 0$ if $B(t)$ does not have
singularities on the positive real axis, and, in particular, if
$A(u,v)>0$.  A semiclassical analysis based on the instanton solutions,
see, e.g., Ref.~\cite{ZJbook}, indicates that the
$\overline{\rm MS}$ and MZM expansions are Borel summable when
\begin{equation}
u + v\geq 0,\qquad N u + v \geq 0.
\label{brr}
\end{equation}
In this region we have 
\begin{equation}
A(u,v) = c \,{\rm Max} \left( u+v,u+\frac{v}{N}\right),
\label{afg}
\end{equation}
where
\begin{equation}
\begin{array}{lcl}
c = \displaystyle{\frac{3}{4}}  & \qquad \qquad & (\overline{\rm MS}),  \\
c = \displaystyle{1.32997\over 8 + N + N^2} & \qquad \qquad & ({\rm MZM}), 
\end{array} 
\label{cval} 
\end{equation}
respectively for the $\overline{\rm MS}$ and MZM expansions.  Under
the additional assumption that the Borel-transform singularities lie
only in the negative axis, the conformal-mapping
method outlined in Refs.~~\cite{LZ-77,ZJbook} turns 
the original expansion into a
convergent one in the region (\ref{brr}).  Alternatively, one may use
the Pad\'e-Borel method, employing Pad\'e approximants to analytically
extend the Borel transform.  

Outside the region (\ref{brr}) the expansion is not Borel summable.
However, for $v>0$, if the condition $u+v >0$ holds, a naive 
extension of the results obtained for the Borel-summable case---but this 
a quite nonrigorous procedure---indicates that the
Borel-transform singularity closest to the origin should still be on
the negative axis. Therefore, the large-order behavior should still be
given by Eq.~(\ref{lobh}) with $A(u,v)$ given by Eq.~(\ref{afg}).
Thus, by using $A(u,v)$ as given by Eq.~(\ref{afg}) and the
conformal-mapping method, one may still take into account the leading
large-order behavior, and therefore hope to get an asymptotic
expansion with a milder behavior, which may still provide reliable
results.  

We should mention that the $\overline{\rm MS}$ series are
essentially four-dimensional, so that they may be affected by
renormalons that make the expansion non-Borel summable for any $u$ and
$v$, and are not detected by the semiclassical analysis
leading to Eqs.~(\ref{lobh}), (\ref{brr}) and (\ref{afg}); see, e.g.,
Ref.~\cite{highorder}.  The same problem should also affect the
$\overline{\rm MS}$ series of O($N$) models.  However, the good
agreement between the results obtained from the FT analyses
\cite{SD-89} and those obtained by other methods indicates that
renormalon effects are either very small or absent (note that, as
shown in Ref.~\cite{BD-84}, this may occur in some renormalization
schemes).  For example, the analysis of the five-loop perturbative
$3d$-$\overline{\rm MS}$ series \cite{SD-89} gives $\nu=0.629(5)$ for the Ising model and
$\nu=0.667(5)$ for the XY model, that are in good agreement with the
most precise estimates obtained by lattice techniques, such as
$\nu=0.63012(16)$ \cite{CPRV-02} and $\nu=0.63020(12)$ \cite{DB-03}
for the Ising model, and $\nu=0.67155(27)$ \cite{CHPRV-01} for the XY
universality class.  On the basis of these results, we will assume
renormalon effects to be negligible in our analyses
of the $3d$-$\overline{\rm MS}$ series.

\subsection{Analysis of the series}
\label{analysisUN}

One can easily identify two FPs
in the theory described by the Lagrangian ${\cal L}_{U(N)}$, without
performing any calculation.  The first one is the Gaussian FP with
$u=v=0$, which is always unstable.  Since, for $v_0=0$, 
${\cal L}_{U(N)}$ is equivalent to the Lagrangian of the
O($N^2+N$)-symmetric vector model, there is an O($N^2+N$) FP
with $v=0$ and $u>0$.  The results of
Ref.~\cite{CPV-03} on the stability of the O($M$)-symmetric FP under
generic perturbations can be used to prove that, for any $N\geq 2$,
the ${\rm O}(N^2+N)$ FP is unstable in theories (\ref{genun}) and
(\ref{gensun}).  Indeed, the term ${\rm Tr} \left( \Phi^\dagger \Phi
\right)^2$ in the Lagrangian ${\cal L}_{U(N)}$, which acts as a
perturbation at the O($N^2+N$) FP, is a particular combination of
quartic operators transforming as the spin-0 and spin-4
representations of the O($N^2+N$) group, and any spin-4 quartic
perturbation is relevant at the three-dimensional O($M$) FP for $M\geq
3$ \cite{CPV-03}, since its RG dimension $y_{4,4}$ is positive for
$M\geq 3$.  In particular, $y_{4,4}\approx 0.27$ at the O(6) FP
\cite{CPV-03}, increases monotonically with increasing $M$
and approaches the value $y_{4,4}=1$ in the large-$M$ limit.

In order to investigate if other FPs are present, one may use the 
$\epsilon$ expansion. A simple analysis, that requires only the one-loop
terms in the expansion of the $\beta$ functions,
indicates that for $\epsilon \ll 1$ only the 
Gaussian and O($N^2+N$) FPs exist and that none of them is stable, 
in agreement with the preceding analysis. Thus, close to four dimensions, the 
transition is of first order. However, there may be FPs that exist 
in three dimensions but are absent for $\epsilon \ll 1$. This is 
indeed what happens in 
the Ginzburg-Landau model of superconductors, in which a complex
scalar field couples to a gauge field \cite{superc}, and in the 
O(2)$\otimes$O($N$) $\Phi^4$ theory describing the critical behavior
of frustrated spin models with noncollinear order
Ref.~\cite{PRV-01,CPPV-04} 
and references therein, and the superfluid transition in
$^3$He \cite{he3}. Thus, to correctly identify the three-dimensional 
critical behavior, we must employ strictly three-dimensional perturbative
schemes. Therefore, we study below the RG flow by using the 
$3d$-$\overline{\rm MS}$ scheme and the MZM scheme. 

For $N=2$
Lagrangian (\ref{genun}) is equivalent to Lagrangian (\ref{LGWch})
written in terms of a real 3$\times$2 matrix field. This model has already been
studied by FT methods \cite{PRV-01,CPPV-04,he3,CPS-02} in the
MZM  (to six loops) and $3d$-$\overline{\rm MS}$ (to five loops) schemes. 
These studies found evidence of two stable FPs: one (called chiral FP) with
attraction domain in the region $g_{2,0}>0$ \cite{PRV-01,CPPV-04} and
another one (called collinear FP) in the region $g_{2,0}<0$ \cite{he3}.  
According to the mapping (\ref{mapg}), the domain $v_0 > 0$ relevant for 
aQCD corresponds to $g_{2,0}<0$, and thus the collinear FP is the relevant
one. Using the results of Ref.~\cite{he3},
we find therefore a stable FP at 
$u^* = -0.54(8)$, $v^* = 1.14(6)$ in the $3d$-$\overline{\rm MS}$ scheme 
and at  $u^* = -3.0(3)$, $v^* = 4.6(3)$ in the MZM scheme.\footnote{
\label{u2xu2}
We mention that a mapping similar to that reported in
Eqs.~(\ref{mapo2xo3}) and (\ref{LGWch}) exists also for 
Lagrangian (\ref{genun}) when $\Phi$ is a generic (not necessarily symmetric)
complex 2$\times$2 matrix. Such a model has a larger symmetry group
$[{\rm U}(2)_L\otimes {\rm U}(2)_R]/{\rm U}(1)_V$ and 
is relevant for two-flavor QCD \cite{PW-84},
when the effect of U(1)$_A$ anomaly is neglected.  Setting $\Phi =
\sum_{\mu=1}^4 \sigma_\mu (\Psi_{\mu 1} + i \Psi_{\mu 2})$, where
$\sigma_\mu\equiv ({\rm Id},\sigma_i)$ and $\Psi$ is a 4$\times$2 real
matrix, Lagrangian (\ref{genun}) corresponds to Lagrangian (\ref{LGWch})
with $g_{1,0} = \frac{3}{2} u_0 + \frac{3}{4}v_0$, $g_{2,0}=-\frac{3}{2}v_0$.
The analysis of the perturbative series in the 
$3d$-$\overline{\rm MS}$ and MZM schemes to five and six 
loops respectively \cite{CPV-04}
shows that theory (\ref{LGWch}) for $N=4$ has a stable FP with
$g_{2}<0$. Therefore, 
the 3-$d$ $[{\rm U}(2)_L\otimes {\rm U}(2)_R]/{\rm U}(1)_V$ model 
has a stable FP, located at $u^*\approx -0.5$, $v^* \approx 1.2$ in the
$3d$-$\overline{\rm MS}$ scheme and in 
$\bar{u}^*=-3.4(3)$ and $\bar{v}^*=5.3(3)$ in the MZM 
scheme (we use here the normalizations of Ref.~\cite{BPV-03}).
Note that this stable FP is not found
close to four dimensions by analyses based on the
$\epsilon$ expansion \cite{PW-84,CP-04-unxum}.  
Its existence implies that systems characterized by
the symmetry-breaking pattern $[{\rm U}(2)_L\otimes {\rm U}(2)_R]/{\rm U}(1)_V 
\rightarrow
{\rm U}_V(2)/{\rm U}_V(1)$ can undergo continuous transitions if they are 
in the attraction domain of this stable FP.
This FP was overlooked in Ref.~\cite{BPV-03}.
Indeed, the analysis of the MZM expansions was limited to the 
region $-2\lesssim \bar{u},\bar{v} \lesssim 4$, that does not include 
the FP reported above.}
These results are confirmed by a direct analysis of the perturbative series in
$u$ and $v$. 

Let us now consider $N= 4$ and investigate the possible existence of 
a stable FP with $v > 0$.
Let us first consider the 
$3d$-$\overline{\rm MS}$ scheme.  In order to find the zeroes of
the $\beta$-functions, we first resum the expansions of $B_u(u,v)$
and $B_v(u,v)$ defined in Eq.~(\ref{Bdef}).  More precisely, we
consider the functions $R_{u,v}(u,v,x)\equiv B_{u,v}(ux,vx)/x^2$.
For each function $R_{u,v}$ we consider several approximants
constructed using the conformal-mapping and Pad\'e-Borel methods.  In
particular, in the case of the conformal-mapping method, we consider
different values of the resummation parameters $\alpha$ and $b$
(typically $\alpha=-1,0,1,2$ and $b=2,\ldots,18$), see
Ref.~\cite{CPV-00} for definitions.


The zeroes of $\beta_{u}$ and $\beta_{v}$ in the region $v\geq 0$
are shown in Fig.~\ref{zeroes4} for $N=4$.
They were obtained by using the conformal-mapping
method and parameters $\alpha=0$ and $b=6$.  
Approximants corresponding to different values of $\alpha$ and $b$, or 
obtained by using the Pad\'e-Borel method, give similar results.
We find no evidence of additional FPs beside the Gaussian and 
the O(20) ones in the $v=0$ axis. 

\FIGURE[ht]{
\epsfig{file=fign4z.eps, width=12truecm} 
\caption{
Zeroes of the $3d$-$\overline{\rm MS}$
$\beta$-functions for $N=4$ in the region $v>0$ (solid
line for $\beta_u$, dashed line for $\beta_v$).  The black dots
indicate the Gaussian and the O(20) FPs.  
The dotted line $4 u + v = 0$ separates the Borel-summable region 
(at the right) from the non-Borel-summable region (at the left).
}
\label{zeroes4}
}

The analysis of the six-loop MZM series was done in a similar way.
We considered the functions $\beta_u(u,v)/u$ and 
$\beta_v(u,v)/v$ and we apply the conformal-mapping method 
as before. The results are 
perfectly consistent with the $3d$-$\overline{\rm MS}$ ones. 
No additional FP is found for $N =  4$.

According to our FT results, a consistent model for a
three-dimensional continuous transition characterized by a complex
symmetric $N\times N$ matrix order parameter and symmetry-breaking
pattern ${\rm U}(N)\rightarrow {\rm O}(N)$ does not exist for 
$N =  4$.  Therefore, the phase transition in such systems must be
of first order. For $N=2$ instead, there is a stable FP; thus, systems
characterized by the symmetry breaking U(2)$\rightarrow$ O(2) can
undergo a continuous transition, if they are in the attraction domain
of the stable FP.

\section{Renormalization-group flow of the {\rm SU}($N$) LGW theory}
\label{rgsun}

We now study the effect of the breaking of ${\rm U}(1)_A$. As we mentioned
in Sec.~\ref{effmod}, we must add to the U($N$) Lagrangian a term proportional
to ${\rm det}\, \Phi + {\rm det}\, \Phi^\dagger$. Such a term is irrelevant
for $N > 4$: in this case the critical behavior is described 
by the U($N$) theory. Therefore, we only consider the cases 
$N=2$ and $N=4$, which are the only ones of interest for 
aQCD. Indeed, $N > 4$ corresponds to $N_f > 2$ and aQCD is asymptotically
free only for $N_f < 11/4 = 2.75$. 

\subsection{The case $N=2$}
\label{n2case}

In the case $N=2$ ($N_f=1$), the determinant is
quadratic in the fundamental field $\Phi$, and other terms must be
added to the effective Lagrangian, i.e.  all possible interactions
with at most four fields. This leads to the LGW Lagrangian
\begin{eqnarray}
&&
{\cal L}_{SU(2)} = {\rm Tr}\, (\partial_\mu \Phi^\dagger) (\partial_\mu \Phi)
+r {\rm Tr}\, \Phi^\dagger \Phi 
+ w_0 \left( {\rm det}\, \Phi^\dagger + {\rm det}\, \Phi \right)
+ {u_0\over 4} \left( {\rm Tr}\, \Phi^\dagger \Phi \right)^2
\nonumber 
\\
&& 
+ {v_0\over 4} {\rm Tr} \left( \Phi^\dagger \Phi \right)^2 
+ {x_0\over 4} \left( {\rm Tr}\, \Phi^\dagger \Phi \right) 
         \left( {\rm det}\, \Phi^\dagger + {\rm det}\, \Phi \right) + 
  {y_0\over 4} \left[ ({\rm det}\, \Phi^\dagger)^2 +
                      ({\rm det}\, \Phi)^2 \right],
\label{su2} 
\end{eqnarray}
where $\Phi_{ij}$ is a complex symmetric 2$\times$2 matrix.  
Note that there is no need to include also a term proportional to 
$({\rm det}\, \Phi)( {\rm det}\, \Phi^\dagger)$. 
Indeed, for any two dimensional 
matrix $M$, we have $2\, {\rm det}\, M = ({\rm Tr}\, M)^2 - {\rm Tr}\, M^2$. 
Therefore, 
\begin{equation}
2 ({\rm det}\, \Phi)( {\rm det}\, \Phi^\dagger) = 
\left( {\rm Tr}\, \Phi^\dagger \Phi \right)^2 - 
{\rm Tr} \left( \Phi^\dagger \Phi \right)^2.
\end{equation}
Since the Lagrangian (\ref{su2}) has two quadratic mass terms, it describes several
transition lines with a multicritical point.

Model (\ref{su2}) can be written in terms of two real
three-component fields $\phi_i$ and $\varphi_i$ by writing
$\Phi=\sum_{j=1}^3 (\phi_j + i \varphi_j) e_j$, where
$(e_1,e_2,e_3)\equiv ({\rm Id}/2,i\sigma_1/2, i\sigma_3/2)$,
${\rm Id}$ is the $2\times 2$ identity matrix, and $\sigma_i$
are the Pauli matrices.  One obtains
\begin{eqnarray}
{\cal L}_{\phi\varphi} &=& {1\over2} (\partial_\mu \phi)^2 + 
             {1\over2} (\partial_\mu \varphi)^2 + 
        {r_\phi\over2} \phi^2 + {r_\varphi\over2} \varphi^2 
\nonumber \\
&&+ {a_0\over 4!} (\phi^2)^2 + {b_0\over 4!} (\varphi^2)^2 + 
     {c_0\over 4!} \phi^2 \varphi^2 + 
     {d_0\over 4!} (\phi \cdot \varphi)^2,
\label{MCR}
\end{eqnarray}
with $r_\phi=r+w_0$, $r_\varphi=r-w_0$, $a_0=\frac{3}{4}(2 u_0 + v_0 +
2 x_0 + y_0)$, $b_0=\frac{3}{4}(2 u_0 + v_0 - 2 x_0 + y_0)$,
$c_0=\frac{3}{2}(2 u_0 + 3 v_0 - y_0)$, and $d_0=-3 (v_0 + y_0)$. 
The mean-field phase diagram of such a model is studied  
in Ref.~\cite{PV-04}. One finds a multicritical point with several 
transition lines. The analysis reported in 
Ref.~\cite{PV-04} indicates that the multicritical point may either belong
to the chiral ${\rm O(2)}\otimes{\rm O(3)}$ universality class 
or may correspond to a first-order transition.
Note that in Ref.~\cite{PV-04} the full RG flow was not studied and 
thus one cannot exclude the presence of other stable FPs, beside the 
${\rm O(2)}\otimes{\rm O(3)}$ chiral FP. This possibility is 
rather unlikely, but an explicit calculation of the RG flow in the 
full parameter space is needed to settle the question. Here, we will 
not consider this additional possibility, since it is of no relevance 
for aQCD, as we discuss below.
In the plane $t$,  $g$, where $g$ parametrizes
the breaking of the symmetry $\phi \leftrightarrow \varphi$, the possible
phase diagrams are reported in Fig.~\ref{figSU2}. 

\FIGURE[ht]{
\epsfig{file=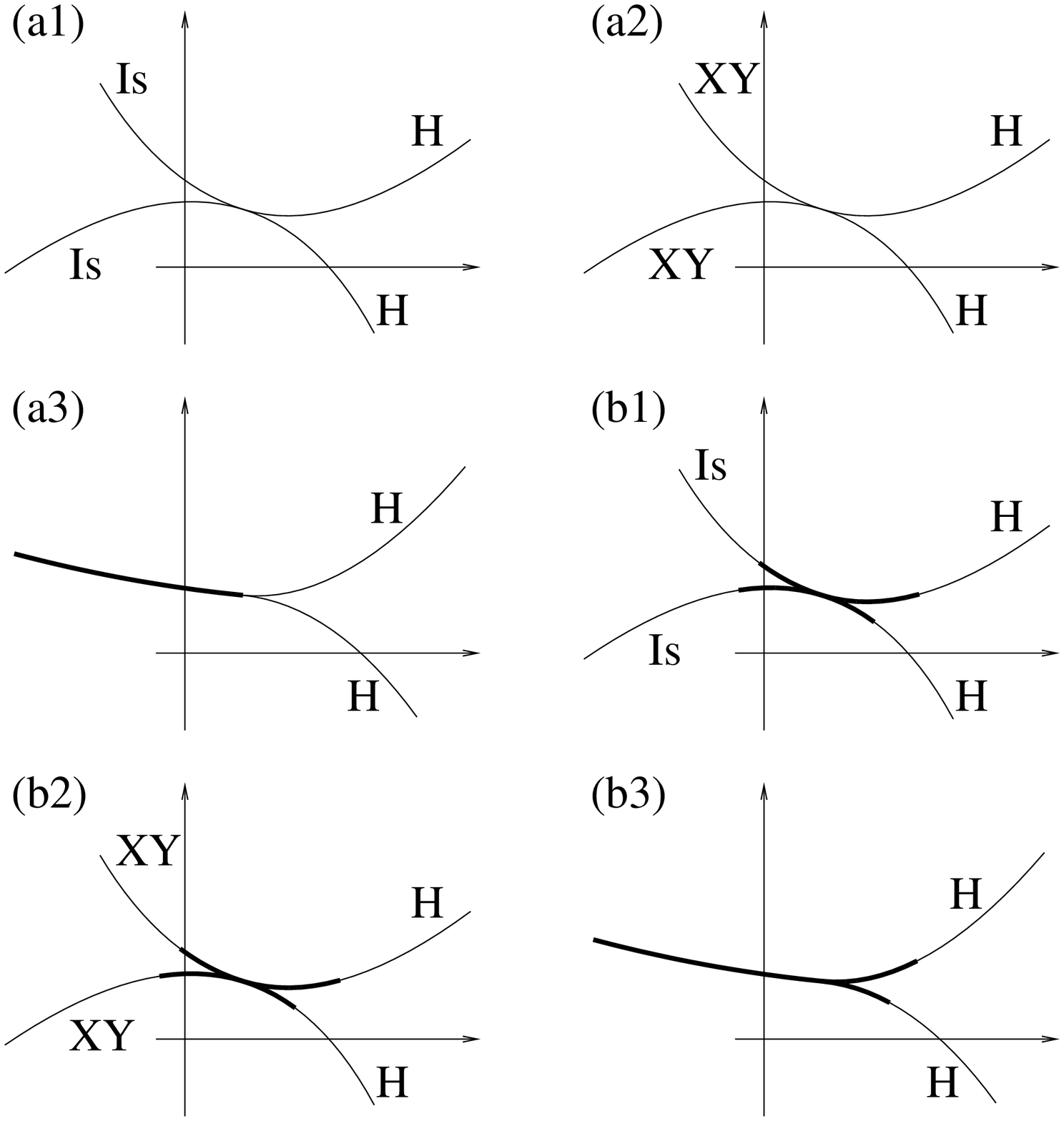, width=12truecm} 
\caption{
Possible phase diagrams for model (\ref{MCR}). In phase diagrams 
(a1), (a2), and (a3) the multicritical point corresponds to a second-order
transition, while in phase diagrams (b1), (b2), and (b3) the 
multicritical transition is
of first order. Thick lines represent first-order transition, 
while thin lines correspond to second-order transitions. We label with 
H, XY, and Is the lines corresponding to transitions belonging to the 
Heisenberg ($N=3$ vector), XY ($N=2$ vector), and Ising 
universality classes, respectively.
}
\label{figSU2}
}

In the case of one-flavor aQCD, the U(2) theory obtained by setting
$w_0=0$, $x_0=0$, $y_0=0$ should correspond to the multicritical point
because of the larger symmetry group.  Therefore, the relevant FP at
the multicritical point is the one found for the U(2) theory in
Sec.~\ref{analysisUN}.  The other FPs of the theory (\ref{MCR}) are
of no relevance since they are not present in the U(2) model.
Thus, if chiral symmetry is not restored at the transition, the
behavior of aQCD as a function of temperature $T$ corresponds to the
behavior observed along a nontrivial line in the $(t,g)$ plane in one
of the phase diagrams reported in Fig.~\ref{figSU2}.~\footnote{ 
A similar scenario with a phase diagram characterized 
by a multicritical point applies to the physically more interesting case of
QCD with two flavors in the fundamental representation.  On the basis
of the results mentioned in footnote \ref{u2xu2}, the presence of a
stable FP in the $[{\rm U}(2)_L\otimes {\rm U}(2)_R]/U(1)_V$ $\Phi^4$
theory (which was overlooked in Ref.~\cite{BPV-03}) has a direct
consequence on the possible phase diagrams in the $T,g$ plane, where
$g$ parametrizes the effective breaking of the U(1)$_A$
symmetry. Indeed, it leaves open the possibility that at the
multicritical point the transition is continuous, with an O(4)
critical line starting from it.  In this case, even an infinitesimal
breaking of the U(1)$_A$ symmetry can give rise to an O(4) critical
behavior. This should partially correct the conclusive remarks
of Ref.~\cite{BPV-03} on the phase diagram relevant for two-flavor
QCD.
}
However, we do
not know which of the phase diagrams applies. Generically, we expect two phase
transitions. One of them, that should occur at higher values of $T$,
may be either of first order or belong to the Heisenberg universality
class [whose symmetry-breaking pattern is SO(3)$\to$SO(2)], which has been
accurately studied in the literature, see, e.g.,
Refs.~\cite{CHPRV-02,PV-r} and references therein. Such a
transition would be associated with the symmetry breaking ${\mathbb
Z}_2 \otimes {\rm SU(2)}\to {\mathbb Z}_2 \otimes {\rm O}(2)$.  The
lower-temperature transition, that may not necessarily exist, may be
of first order, or continuous.  According to Fig.~\ref{figSU2}, the
transition may belong either to the Ising or to the XY universality
class. It is easy to see that in the latter case the U(1)$_V$ symmetry
would be broken, violating the Vafa-Witten theorem
\cite{VW-84}. Therefore, the only possibility is an Ising transition
that corresponds to the breaking ${\mathbb Z}_2 \otimes {\rm O}(2) \to
{\mathbb Z}_2 \otimes {\rm SO}(2)$.  Note that the symmetry breaking
${\rm SU}(2) \to {\rm SO}(2)$ is realized here through two different
transitions and that an SU(2)/SO(2) universality class does 
not exist. One may discriminate between the one- or two-transition
hypothesis by determining the symmetry of the $T=0$ phase, which 
differs by a  ${\mathbb Z}_2$ group in the two cases.

\subsection{The case $N=4$}
\label{n4case}

For $N=4$ ($N_f=2$) the determinant is a quartic-order term, giving
rise to a generalized LGW $\Phi^4$ theory (\ref{gensun}) with three
quartic parameters.  In this case stability requires 
\begin{equation}
u_0+v_0 \geq 0,\qquad 4 u_0+ v_0 - 2|w_0|\geq 0.
\end{equation}
As discussed in the Appendix, the symmetry-breaking pattern and vacuum
structure appropriate for aQCD are realized for 
$v_0 > - {3\over2} |w_0|$. 
Note that the FPs already identified in the
${\rm U}(4)$ theory, i.e. the Gaussian and the 
O(20) FP,  are also FPs of the ${\rm SU}(4)$ theory.
They are both unstable and thus are of no relevance for the critical 
behavior. 

In order to study the RG flow of the theory (\ref{gensun}) for $N=4$,
we consider the $3d$-$\overline{\rm MS}$ scheme and the MZM scheme as 
before. In the $3d$-$\overline{\rm MS}$ scheme we set
\begin{eqnarray}
\Phi &=& [Z_\phi(u,v,w)]^{1/2} \Phi_R, \\
u_0 &=& A_d \mu^\epsilon Z_u(u,v,w) , \nonumber \\
v_0 &=& A_d \mu^\epsilon Z_v(u,v,w) , \nonumber \\
w_0 &=& {1\over 24} A_d \mu^\epsilon Z_w(u,v,w) , \nonumber 
\end{eqnarray}
and determine the corresponding $\beta$ functions to five loops. They
are given by
\begin{eqnarray}
\beta_u &=& -\epsilon u + 7 u^2 + 5 u v +  \frac{3}{4} v^2 + 3 w^2 -
\frac{111}{8} u^3 - \frac{55}{4} u^2 v -  \frac{29}{4} uv^2 
\label{bbu}\\
&&- \frac{9}{4}v^3 
- \frac{33}{2} u w^2 - 3 v w^2 +
\sum_{i+j+k\geq 4} c^{u}_{ijk} u^i v^j w^k,
\nonumber
\\
\beta_v &=& -\epsilon  v + 3 u v +  \frac{13}{4} v^2 - 3 w^2 -
 \frac{91}{8} u^2 v -  16 uv^2 - 6 v^3 
\label{bbv}\\
&&+ 9 u w^2 + \frac{9}{2} v w^2 +
 \sum_{i+j+k\geq 4}  c^{v}_{ijk} u^i v^j w^k, 
\nonumber\\
\beta_w &=& -\epsilon  w + 3 u w - \frac{3}{2} v w  
- \frac{91}{8} u^2 w -\frac{7}{4} v^2w - \frac{7}{4} u v w
\label{bbw}\\
&&+ \frac{3}{2} w^3 
+ \sum_{i+j+k\geq 4}  c^{w}_{ijk} u^i v^j w^k.
\nonumber
\end{eqnarray}
The coefficients $c_{ijk}$ up to five loops, i.e.
for $4\leq i+j+k\leq 6$, are reported in Table~\ref{tabc}.
We also computed the five-loop expansion of the
RG functions $\eta_{\phi,t}$ defined as in Eq.~(\ref{etadefmsb}):
\begin{equation}
\eta_{\phi,t} =  \sum_{i+j+k\geq 1} e^{\phi,t}_{ijk} u^i v^j w^k\; .
\label{etaphit}
\end{equation}
The coefficients $e^\phi_{ijk}$ and  $e^t_{ijk}$
up to five loops, i.e.
for $1\leq i+j+k\leq 5$, are reported in Table~\ref{tabeta}. 
The standard critical exponents are related to $\eta_{\phi,t}$ by
\begin{equation}
\eta = \eta_\phi(u^*,v^*,w^*),\qquad
\nu = \left[ 2 + \eta_t(u^*,v^*,w^*) - \eta_\phi(u^*,v^*,w^*) \right] ^{-1},
\label{exponents2} 
\end{equation}
where $u^*$, $v^*$, $w^*$ are the coordinates of the stable FP in the
case it exists.

In the MZM scheme,
beside the renormalization conditions (\ref{ren1}) and
(\ref{ren2}), we also set
\begin{eqnarray}
\hat{\Gamma}^{(4)}_{a_1a_2,b_1b_2,c_1c_2,d_1d_2}(0) &=&  
\left. {\delta^4 \Gamma \over \delta \Phi_{a_1a_2}
\delta \Phi_{b_1b_2} \delta \Phi_{c_1c_2} \delta \Phi_{d_1d_2}}\right|_{\rm zero \; mom.}
\nonumber \\
&=& Z_\phi^{-2} m^{4-d} {\pi\over 21}
w \epsilon_{a_1b_1c_1d_1} \epsilon_{a_2b_2c_2d_2} ,
\label{ren3}  
\end{eqnarray}
where $\epsilon_{ijkl}$ is the completely antisymmetric tensor ($\epsilon_{1234}=1$).
We computed six-loop series in the MZM scheme.
The MZM $\beta$ functions are given by
\begin{eqnarray}
\beta_u &=& - u + u^2 + \frac{5}{7} u v +  \frac{3}{28} v^2 +  \frac{3}{7} w^2 
+\sum_{i+j+k\geq 3} d^{u}_{ijk} u^i v^j w^k,
\label{bbug}\\
\beta_v &=& - v + \frac{3}{7} u v +  \frac{13}{28} v^2 - \frac{3}{7} w^2 
+ \sum_{i+j+k\geq 3}  d^{v}_{ijk} u^i v^j w^k, 
\label{bbvg}\\
\beta_w &=& - w + \frac{3}{7} u w - \frac{3}{14} v w  
+ \sum_{i+j+k\geq 3}  d^{w}_{ijk} u^i v^j w^k.
\label{bbwg}
\end{eqnarray}
The coefficients $d_{ijk}$ up to six loops, i.e.
for $3\leq i+j+k\leq 7$, are reported in Table~\ref{tabcg}.
We also computed the RG functions $\eta_{\phi,t}$
defined in Eq.~(\ref{etadefmzm}) to six loops.
The coefficients $e^\phi_{ijk}$ and  $e^t_{ijk}$
defined in Eq.~(\ref{etaphit})
are reported in Table~\ref{tabeta}.

Again, close to four dimensions
there are only two FPs, the Gaussian and the O(20) FPs
(the latter located at $u=\epsilon/7$, $v=0$, and $w = 0$), 
i.e. those already found in the U(4) theory, which are both unstable.
In order to investigate the possible existence of other FPs in three 
dimensions, we analyze the $3d$-$\overline{\rm MS}$ and MZM 
perturbative expansions. 
Note that the model is invariant under the transformations 
$\Phi \to e^{i \pi/4} \Phi$, $(u,v,w) \to (u,v,-w)$. This implies that 
$\beta_u$, $\beta_v$, and $\beta_w/w$ are even in $w$. Therefore, we can
restrict our search of FPs to the $w>0$ space; if a FP 
with coordinates $u,v,w>0$ exists, there is also another FP with the same 
critical properties at $u,v,-w$.  

We use the conformal-mapping method 
already employed in Sec.~\ref{analysisUN} and the large-order behavior 
of the perturbative series.
Writing
\begin{equation}
S(x u, x v, x w) = \sum_n s_n(u,v,w) x^n,
\end{equation}
semiclassical calculations based on instanton solutions give
\begin{eqnarray}
&&s_n \sim n! [-A(u, v, w)]^n \, n^b,
\label{bsingw}\\
&&A(u,v,w) = c \,
{\rm Max} \left( u+v,\, 
u+\frac{v}{4} + \frac{|w|}{2}\right),
\nonumber
\end{eqnarray}
where the constant $c$ is given by
$c=3/4$ and $c=0.0474989$ respectively for the
$\overline{\rm MS}$ and MZM expansions.  
The perturbative expansions are Borel summable for
\begin{equation}
u+v \geq 0,\qquad 4 u+v - 2 |w|\geq 0 \; .
\label{n4sunbr}
\end{equation}

The $3d$-$\overline{\rm MS}$ five-loop series are analyzed as in the 
U($N$) case and indicate the presence of a new FP with $w\not = 0$. 
Most approximants of the five-loop series
(more than 90\% of the approximants with $\alpha=-1,0,1,2$ and $b=3,\ldots,18$)
present a common zero with $w > 0$.
They are approximately 50\% at four loops.
This new FP is
located at $u^*=0.23(9)$, $v^*=0.34(6)$, and $w^*=0.54(8)$, where the
error takes into account the spread of the results of the approximants
considered. Note that the error (spread of the
results) is rather large, essentially because this FP lies relatively
far from the origin (note that the O(20) FP lies at
$u^*\approx 0.17$), and also because it is close to the boundary of
the Borel summable region.  The analysis of the corresponding
stability matrix shows that this FP is stable, and therefore it determines 
the universal properties of 
continuous transitions in systems described by the SU(4) LGW
$\Phi^4$ theory (\ref{gensun}).  We also computed the corresponding
critical exponents by evaluating the RG functions $\eta_{\phi,t}$ at
the FP. 
From the analysis of the expansions of
$\eta_\phi$ and $(2+\eta_t-\eta_\phi)^{-1}$, we obtain $\eta=0.23(8)$
and $\nu=1.1(3)$, where the error takes into account the spread of the
approximants and the uncertainty on the FP coordinates.  
The presence of the stable FP is confirmed by an analysis
based on Pad\'e-Borel approximants; indeed, using [4/1] Pad\'e
approximants for all $\beta$ functions, one obtains $u^*\approx
0.29(5)$, $v^*\approx 0.23(1)$, and $w^*\approx 0.51(4)$ (the errors
indicate how the results change when the parameter $b$ is varied 
between 4 and 18), 
which is substantially consistent with the results
of the conformal-mapping analysis. Apparently, the Pad\'e-Borel 
method gives a more accurate estimate of the location of the FP.
But, note that the error is only related to the dependence
on $b$ of a single Pad\'e approximant; thus, it is likely to be 
underestimated. As we have already seen in many 
other instances, the conformal-method error should provide a more 
realistic estimate of the real accuracy of the result.

The rather low precision of the $3d$-$\overline{\rm
MS}$ results may give rise to some doubts on the existence itself
of the stable FP, and therefore it calls for an independent and
nontrivial crosscheck.  This is provided by the analysis of the
perturbative expansions in the alternative three-dimensional MZM scheme. 
We follow the same steps as in the U($N$) case, considering 
the conformal-mapping  method and approximants with 
$\alpha=-1,0,1,2$ and $b=3,...18$ (see Ref.~\cite{CPV-00} for definitions).
The analysis confirms the presence of a stable FP
at $u^*=0.0(3)$, $v^*=3.5(2)$ and
$w^*=4.6(2)$, where the error is related to the
spread of the results.  The presence of a 
FP is stable with respect to the number of loops.  Indeed, at 
five loops (resp. four loops) we find a FP
at $u^*=0.4(2)$, $v^*=3.2(2)$ and $w^*=4.3(2)$ (resp.
$u^*=0.7(3)$, $v^*=3.6(5)$ and $w^*=4.6(6)$). For              
comparison, in this scheme the O(20) FP is located at $u^*\approx 1.18$
\cite{AS-95}.  We also estimate the critical exponents, finding
$\eta=0.20(3)$ and $\nu=1.1(3)$ (again from the analysis of the expansions
of $\eta_\phi$ and $(2+\eta_t-\eta_\phi)^{-1}$), in substantial
agreement with the results of the $3d$-$\overline{\rm MS}$ scheme.
Also in this scheme results are not very precise, 
although they are  apparently more accurate 
than the $3d$-$\overline{\rm MS}$ ones.

In conclusion, the analysis of the perturbative expansions in the
3$d$-$\overline{\rm MS}$ and MZM schemes provides evidence for the
existence of a stable FP, and therefore of a new universality class
that describes continuous transitions in systems described by the LGW
Lagrangian (\ref{gensun}) with $N=4$ and symmetry-breaking pattern
${\rm SU(4)} \to {\rm SO(4)}$.  Although the results of the analyses
do not appear particularly precise, the
substantial agreement between the two schemes makes us confident on
their reliability.  Note that the existence of a FP does
not exclude the possibility of observing first-order transitions; indeed,
this is still possible for systems outside the attraction domain of
the stable FP.  

We recall that in Sec.~\ref{analysisUN} we found no stable FP in the
U(4) LGW $\Phi^4$ theory, which should be relevant if the U(1)
symmetry broken by the anomaly is effectively restored at $T_c$, thus
suggesting first-order transitions for the corresponding systems.  The
existence of a stable FP with $w\not=0$ in the SU(4) theory is an
interesting example of the so-called phenomenon of softening of
first-order transitions: by breaking some symmetry of the original
model, a first-order transition may become a second-order one. This
phenomenon is well known in spin systems: the introduction of quenched
disorder may soften the first-order transitions of pure systems
\cite{softening}. In disordered systems translational invariance is
the broken symmetry. In our case, it is instead the internal symmetry
of the system that gets reduced.

\acknowledgments{
We thank Maurizio Davini for his valuable technical support.}

\appendix
\section{Vacua of the U($N$) and SU(4) theories} \label{AppA}

Let us consider the potential
\begin{equation}
V_{U(N)}(\Phi) = r {\rm Tr}\, \Phi^\dagger \Phi
+ {u_0\over 4} \left( {\rm Tr}\, \Phi^\dagger \Phi \right)^2
+ {v_0\over 4} {\rm Tr} \left( \Phi^\dagger \Phi \right)^2 ,
\end{equation}
and let us determine the fields that minimize it. 
The stationarity condition gives the equation
\begin{equation}  
r \Phi_{ab} + {u_0\over 2} \Phi_{ab} {\rm Tr}\, \Phi^\dagger \Phi + 
  {v_0\over2} \left( \Phi\Phi^\dagger\Phi\right)_{ab} = 0.
\label{eq-station}
\end{equation}
If $\Phi$ is a solution of Eq.~(\ref{eq-station}), it is easy to show that 
\begin{equation}
\left. V_{U(N)}(\Phi)\right|_{\rm sol} =  
      {r\over2} {\rm Tr}\, \Phi^\dagger \Phi.
\label{Vsol}
\end{equation}
In order to determine the solutions of Eq.~(\ref{eq-station}), let us 
perform a polar decomposition of $\Phi$: we write $\Phi = P U $,
where $P = \sqrt{\Phi^\dagger \Phi}$ is a hermitian positive semidefinite
matrix and $U$ is a unitary matrix. Then we diagonalize $P$, writing 
$P = V P_d V^\dagger$ with 
$P_d = {\rm diag}(\lambda_1,\lambda_2,\ldots,\lambda_N)$, 
$\lambda_1 \ge\lambda_2 \ge \ldots \ge \lambda_N\ge 0$, and 
$V$ unitary.
Now, if $k$ is the rank of $\Phi$, we have for $1 \le a \le k$
the equation
\begin{equation}
r + {u_0\over 2} {\rm Tr}\, P^2_d + 
   {v_0\over2} \lambda_a^2 = 0,
\label{eq2}
\end{equation}
that shows that $\lambda_a$ does not depend on $a$.
Summing this equation over $1 \le a \le k$, we obtain
\begin{equation}
{\rm Tr}\, \Phi^\dagger \Phi  = 
{\rm Tr}\, P^2_d = \sum_{a=1}^k \lambda_a^2 = 
       - {2 k r\over k u_0 + v_0}.
\label{trace}
\end{equation}
Since $u_0 + v_0 > 0$ and $N u_0 + v_0 > 0$, for $r > 0$ 
we must have $k = 0$, i.e. the only possible solution is $\Phi  = 0$:
for $r > 0$ the system is disordered. For $r < 0$ instead, any $k$ with
$0\le k \le N$ is acceptable. Using Eq.~(\ref{Vsol}) we obtain
\begin{equation}
\left. V(\Phi)\right|_{\rm sol} =  - {k r^2\over k u_0 + v_0},
\end{equation}
which shows that the energy depends only on the rank $k$ of the solution.
A simple calculation shows that the minimum of the potential is attained
for $k = N$ if $v_0 > 0$ and $k = 1$ if $v_0 < 0$.

For $k = N$, we have
\begin{equation}
P = - {2r\over N u_0 + v_0}\, {\rm Id},
\end{equation}
where $\rm Id$ is the identity matrix.
Thus, $\Phi \propto  Q$, where $Q$ is a symmetric unitary 
matrix. Of course, being U($N$) connected, we can write 
$Q = U^2 = U U^T$, with $U$ unitary and symmetric.
Thus, modulo symmetry transformations, we may take 
$\Phi_{\rm min} \propto {\rm Id}$.

If $k = 1$, $\Phi_{ab} = v_a v_b$, where $v_a$ is an $N$-dimensional 
complex vector. Since, for any $v$ there exists $U\in {\rm U}(N)$ such that 
$v_a = \sum_{b} U_{ab} w_b$, where $w = (a,0,\ldots,0)$, with $a$ real
and positive, we can write $\Phi$ as 
\begin{equation}
\Phi = - {2 r\over u_0 + v_0} U I_1 U^T,
\end{equation}
with $I_1 = {\rm diag}(1,0,\ldots,0)$. Thus, modulo symmetry transformations, 
we may take $\Phi_{\rm min} \propto {\rm I_1}$. 

Let us now determine the symmetry-breaking patterns. The theory is invariant 
under the transformations $\Phi\to U \Phi U^T$, $U$ unitary, which form 
a group isomorphic to U($N)/{\mathbb Z}_2$ (the quotient is due to the 
fact that $U$ and $-U$ give rise to the same transformation). 
The transformations that leave invariant $\Phi_{\rm min} \propto {\rm Id}$
are those with $U\in {\rm O}(N)$, so that the invariance group is 
${\rm O}(N)/{\mathbb Z}_2$. If $\Phi_{\rm min} \propto {\rm I_1}$, 
the transformations that leave invariant $\Phi_{\rm min}$ have the form 
$U = (\pm 1)\oplus R$, $R\in {\rm U}(N-1)$, so that the invariance 
group is $[{\rm U}(N-1)\otimes {\mathbb Z}_2]/{\mathbb Z}_2 \cong
{\rm U}(N-1)$. Beside $\Phi\to U \Phi U^T$, the model, ${\rm Id}$, and $I_1$
are invariant under the ${\mathbb Z}_2$ transformations 
$\Phi \to \Phi^\dagger$. Thus, the symmetry breaking pattern is 
${\rm U}(N) \to {\rm O}(N)$  for $v_0 > 0$ and 
${\rm U}(N) \to {\rm U}(N-1)\otimes {\mathbb Z}_2$ for $v_0 < 0$.
Note that for $N=2$ they are identical since 
${\rm O}(2) \cong {\rm SO}(2) \otimes {\mathbb Z}_2 \cong 
{\rm U}(1) \otimes {\mathbb Z}_2$. 

To conclude, let us note that, as a consequence of the 
 Vafa-Witten theorem \cite{VW-84}, 
the vector group $U(N_f)_V$ can never be broken. Therefore, the 
vacuum must be invariant under the $U(N/2)$ transformations
\begin{equation}
U_V \equiv \left( 
   \begin{array}{cc}
    U & 0 \\ 0 & U^* 
   \end{array} \right),
\end{equation}
with $U\in U(N/2)$. It is easy to verify that\footnote{
Indeed, if we write 
\begin{equation}
Q = \left(
   \begin{array}{cc}
    a  & b \\ b^T & c^*
   \end{array} \right)\; ,
\end{equation}
where $a$, $b$, and $c$ are $N/2\times N/2$ matrices, 
we must have $UaU^T = a$, $Uc U^T = c $, $UbU^\dagger = b$. 
The first two conditions imply $a = c = 0$ (it is enough to consider 
infinitesimal matrices $U$), while the third one gives $b \propto {\rm Id}$.
}
the only symmetric matrices $Q$ such that $U_V Q U_V^T = Q$ are proportional to
\begin{equation}
J \equiv \left(
   \begin{array}{cc}
    0  & {\rm Id} \\ {\rm Id}  & 0
   \end{array} \right) \; ,
\end{equation}
where ${\rm Id}$ is the $N/2\times N/2$ identity matrix.
Thus, the only possible vacuum for aQCD is proportional to $J$. It is easy to 
verify that this is possible for $v > 0$ (as we showed any matrix proportional
to a symmetric unitary 
matrix is a possible vacuum), but not for $v < 0$. Thus, in the aQCD case we 
must restrict ourselves to the case $v > 0$. 

Now let us consider the effect of the determinant for $N=4$. The 
potential is 
\begin{equation}
V_{SU(N)} (\Phi)= V_{U(N)}(\Phi)
 + w_0 ({\rm det}\, \Phi + {\rm det}\, \Phi^\dagger).
\end{equation}
The stationarity condition gives then
\begin{equation}
r\Phi_{ab} + {u_0\over2} \Phi_{ab} {\rm Tr}\, \Phi^\dagger \Phi + 
   {v_0\over2} (\Phi\Phi^\dagger \Phi)_{ab} + 
   {w_0\over6} \sum_{c_1 d_1 e_1 c_2 d_2 e_2} 
   \epsilon_{bc_1d_1e_1} \epsilon_{ac_2d_2e_2}
   \Phi_{c_1c_2}^\dagger \Phi_{d_1d_2}^\dagger \Phi_{e_1e_2}^\dagger = 0.
\end{equation}
Now, we write again $\Phi = V P_d V^\dagger U = 
\hat{V} P_d [({\rm det}\, V)^{1/2} V^\dagger U V^*] \hat{V}^T$ with $P_d$ 
diagonal and $\hat{V} = V({\rm det}\, V)^{-1/4}$. 
Thus, modulo a symmetry transformation,
we can simply write $\Phi = P_d U$, with $U$ unitary. 
Substituting in the previous equation we obtain 
the equation
\begin{equation}
r\lambda_a + {u_0\over2} \lambda_a {\rm Tr}\, P_d^2 +
     {v_0\over2} \lambda_a^3  - 
    w_0 ({\rm det}\, U)  \prod_{c\not=a} \lambda_c = 0.
\label{eqgenSU4}
\end{equation}
The term proportional to $w_0$ is the product of three eigenvalues
and thus it is relevant only if the rank of $\Phi$ is 3 or 4. 
If the rank is three, assume $\lambda_4 = 0$. The previous equation for 
$a = 4$ gives 
\begin{equation}
    \lambda_1 \lambda_2 \lambda_3 = 0,
\end{equation}
that contradicts the hypothesis $\lambda_1 \lambda_2 \lambda_3\not=0$.
Thus, there is no solution with rank $k = 3$.

Assume now that the rank $k$ is  4. 
Since all $\lambda_a$ are real and nonvanishing, $({\rm det}\, U)$
must be real, hence equal to $s = \pm 1$, since $U$ is unitary. 
Let us now 
determine the eigenvalues. Eq.~(\ref{eqgenSU4}) implies 
\begin{equation}
R_{ab} \equiv \lambda_a E_a - \lambda_b E_b = 
  (\lambda_a^2 - \lambda_b^2) 
  \left[r + {u_0\over2} {\rm Tr}\, P_d^2 + 
        {v_0\over2} (\lambda_a^2 + \lambda_b^2)
   \right] = 0.
\label{eqtraccia}
\end{equation} 
Now, consider $R_{12} = 0$, $R_{13} = 0$, and $R_{23} = 0$. It is immediate 
to verify that these equations imply that at least two eigenvalues among
$\lambda_1$, $\lambda_2$, $\lambda_3$ are equal. Without loss of generality
we assume $\lambda_1 = \lambda_2$. An analogous discussion indicates that
at least two eigenvalues among $\lambda_1$, $\lambda_3$, $\lambda_4$ are equal. 
Thus, there are two possible cases: $\lambda_1 = \lambda_2 = \lambda_3$;
$\lambda_1 = \lambda_2$ and $\lambda_3 = \lambda_4$. 
Finally, consider $R_{24} = 0$; we obtain that either $\lambda_2 = \lambda_4$
or $2 r + {u_0} {\rm Tr}\, P_d^2 +
        {v_0} (\lambda_2^2 + \lambda_4^2) = 0$, that gives a relation 
between $\lambda_2$ and $\lambda_4$. In conclusion, we obtain 
two different classes of solutions: 
\begin{itemize}
\item[(i)] all eigenvalues are equal with
\begin{equation}
\lambda^2_a = {- 2r\over 4 u_0 + v_0 + 2 s w_0}
\label{solSU4}
\end{equation}
and potential
\begin{equation}
\left. V(\Phi)\right|_{\rm sol} = 
  {- 4 r^2
            \over 4 u_0 + v_0 + 2 s w_0};
\label{potsolSU4}
\end{equation}
\item[(ii)] one eigenvalue differs from the others:
\begin{eqnarray}
&& \lambda_1^2 = \lambda_2^2 = \lambda_3^2  = 
      {-2 r v_0^2\over 
       3 u_0 v_0^2 + v_0^3 + 4 u_0 w_0^2 + 4 v_0 w_0^2}, \nonumber \\
&& \lambda_4 = {2 s w_0 \over v_0} \lambda_1, 
\end{eqnarray}
with potential 
\begin{equation}
\left. V(\Phi)\right|_{\rm sol} =
- r^2 {3 v_0^2 + 4 w_0^2 \over 
      3 u_0 v_0^2 + v_0^3 + 4 u_0 w_0^2 + 4 v_0 w_0^2}.
\end{equation}
Note that the solution with $s = 1$ exists only if $w_0 v_0$ is positive, 
the one with $s = -1$ in the opposite case. For $w_0 = 0$ this solution
corresponds to the rank-3 solution we have determined before.
\end{itemize}

Comparing the value of the potential for the different solutions, we obtain 
finally:
\begin{itemize}
\item[(a)] for $v_0 < 0$ and $|w_0| < - 2 v_0/3$ the relevant solution
has rank $k = 1$ and thus, modulo symmetry transformations, we can take 
$\Phi_{\rm min} \propto I_1$. 
The symmetry of the vacuum is SU(3)$\otimes {\mathbb Z}_2$. 
Such a case is not of interest for aQCD, since it is not invariant under
U(2)$_V$. 
\item[(b)]
for $w_0 > 0$ and $w_0 > - 2 v_0/3$ the relevant solution is given 
by Eq.~(\ref{solSU4}) with potential (\ref{potsolSU4}), setting 
$s = -1$. Therefore, $U$ is symmetric with 
${\rm det}\, U = -1$. It follows that $U = e^{i\pi/4} V V^T$, 
$V \in {\rm SU}(4)$.
Modulo symmetry transformations, one can therefore choose 
$\Phi_{\rm min}\propto e^{i\pi/4} {\rm Id}$.
The vacuum is invariant under ${\rm SO(4)}$ (note that the 
symmetry $\Phi \to \Phi^\dagger$ is broken here). 
This case is not relevant for aQCD since the vacuum breaks 
$\Phi \to \Phi^\dagger$. 

\item[(c)]
for $w_0 < 0$ and $|w_0| > - 2 v_0/3$ the relevant solution is given 
in Eq.~(\ref{solSU4}) with potential (\ref{potsolSU4}), setting 
$s = 1$. Therefore, $U$ is symmetric with 
${\rm det}\, U = 1$. It follows that $U = V V^T$ with $V\in $ SU(4).
Modulo symmetry transformations, one can therefore choose 
$\Phi_{\rm min}\propto {\rm Id}$.
The vacuum is invariant under ${\rm SO(4)}\otimes {\mathbb Z}_2$. 
The matrix $J$ is one of the possible vacuum solutions and thus this
case is of relevance for aQCD.
\end{itemize}

\clearpage

\TABLE[ht]{
\footnotesize
\caption{$3d$-$\overline{\rm MS}$ scheme: 
coefficients $a^{u}_{ij}$, cf. Eq.~(\ref{bun}).  
}
\label{betau}
\begin{tabular}{cl}
\hline\hline
\multicolumn{1}{c}{$i,j$}&
\multicolumn{1}{c}{$a^u_{ij}$}\\
\hline \hline
4,0   &   $10.7397 + 2.92770 N + 2.99216 N^2 + 0.128906 N^3 + 0.0644531 N^4$\\
3,1   &   $17.5092 + 18.1264 N + 1.23437 N^2 + 0.617187 N^3$\\
2,2   &   $25.2599 + 9.50698 N + 3.95252 N^2 + 0.0117187 N^3 + 0.00292968 N^4$\\
1,3   &   $13.4014 + 8.45267 N + 0.626953 N^2 + 0.117187 N^3$\\
0,4   &   $3.61949 + 1.54627 N + 0.301169 N^2$\\
\hline
5,0   &   $-63.5743 - 21.0606 N - 22.3638 N^2 - 2.60592 N^3 - 1.30194 N^4 $\\ 
& $+0.00122070 N^5 + 0.000406901 N^6$\\
4,1   &   $ -140.774 - 153.387 N - 25.2352 N^2 - 12.6425 N^3 - 0.0298321 N^4 $\\ 
& $- 0.00994404 N^5$\\
3,2   &   $-260.744 - 143.388 N - 66.5772 N^2 - 2.01164 N^3 - 0.526260 N^4$\\
2,3   &   $ -212.910 - 157.262 N - 23.2253 N^2 - 5.14407 N^3 - 0.0223761 N^4 $\\ 
&$- 0.00344248 N^5$\\
1,4   &   $-107.874 - 60.8991 N - 15.9106 N^2 - 0.709064 N^3 - 0.0784548 N^4$\\
0,5   &   $-20.6853 - 12.8827 N - 2.21799 N^2 - 0.215929 N^3$\\
\hline
6,0   &   $475.552 + 183.011 N + 200.1526 N^2 + 34.5561 N^3 + 17.9613 N^4 + 0.818270 N^5 $\\ 
& $+ 0.269906 N^6 - 0.00244287 N^7 - 0.000610718 N^8$\\
5,1   &   $1339.84 + 1524.75 N + 373.269 N^2 + 195.251 N^3 + 10.3200 N^4 + 3.40600 N^5 $\\ 
& $- 0.0291545 N^6 - 0.00728864 N^7$\\
4,2   &   $ 3025.04 + 2080.71 N + 1065.77 N^2 + 97.1136 N^3 + 29.0928 N^4 + 0.178866 N^5 $\\ 
& $+ 0.0283466 N^6$\\
3,3   &   $3355.06 + 2824.44 N + 654.568 N^2 + 172.564 N^3 + 4.47104 N^4 + 0.769792 N^5$\\
2,4   &   $2471.77 + 1715.81 N + 548.534 N^2 + 48.2367 N^3 + 6.77586 N^4 + 0.0200149 N^5 $\\ 
& $+ 0.00195671 N^6$\\
1,5   &   $944.177 + 692.287 N + 168.563 N^2 + 25.2091 N^3 + 0.760372 N^4 + 0.0526262 N^5$\\
0,6   &   $157.642 + 109.278 N + 29.0944 N^2 + 2.78840 N^3 + 0.163781 N^4$\\
\hline\hline
\end{tabular}
}

\TABLE[ht]{
\footnotesize
\caption{$3d$-$\overline{\rm MS}$ scheme: 
coefficients $a^{v}_{ij}$, cf. Eq.~(\ref{bvn}).  
}
\label{betav}
\begin{tabular}{cl}
\hline\hline
\multicolumn{1}{c}{$i,j$}&
\multicolumn{1}{c}{$a^v_{ij}$}\\
\hline \hline
4,0   &   $0$\\
3,1   &   $25.4497 + 2.33904 N + 2.288 N^2 - 0.101562 N^3 - 0.0507812 N^4 $\\
2,2   &   $39.1784 + 23.3634 N - 0.0214843 N^2 - 0.136718 N^3 $\\
1,3   &   $29.5574 + 12.9326 N + 2.32341 N^2 $\\
0,4   &   $7.12024 + 3.75368 N + 0.486716 N^2 + 0.0253906 N^3 $\\
\hline
5,0   &   $0$\\
4,1   &   $ -177.097 - 22.3032 N - 22.5027 N^2 - 0.392090 N^3 - 0.178529 N^4 $\\ 
&$+ 0.0210195 N^5 + 0.00700652 N^6 $\\
3,2   &   $ -374.999 - 241.790 N - 14.5218 N^2 - 4.72881 N^3 + 0.179072 N^4 
+ 0.0584779 N^5 $\\
2,3   &   $ -422.833 - 232.645 N - 55.3180 N^2 + 0.1763229 N^3 + 0.137500 N^4 $\\
1,4   &   $ -209.997 - 132.747 N - 23.9377 N^2 - 2.37050 N^3 $\\
0,5   &   $ -42.8890 - 25.8808 N - 5.72264 N^2 - 0.396930 N^3 - 0.0136094 N^4 $\\
\hline
6,0   &   $0$\\
5,1   &   $1513.47 + 243.234 N + 250.177 N^2 + 13.8095 N^3 + 6.71436 N^4 $\\ 
&$- 0.228540 N^5 - 0.0762733 N^6 - 0.0000799564 N^7 - 0.0000199891 N^8 $\\
4,2   &   $4108.25 + 2830.72 N + 327.369 N^2 + 121.934 N^3 - 0.955058 N^4 $\\ 
&$- 0.471074 N^5 - 0.00569730 N^6 - 0.00109038 N^7 $\\
3,3   &   $6155.99 + 3929.91 N + 1110.53 N^2 + 33.3855 N^3 + 4.60723 N^4 $\\ 
&$+ 0.0693464 N^5 + 0.00888516 N^6 $\\
2,4   &   $4661.51 + 3367.08 N + 766.260 N^2 + 98.8345 N^3 - 0.0328212 N^4 
- 0.0247545 N^5 $\\
1,5   &   $1909.14 + 1333.31 N + 360.926 N^2 + 37.1082 N^3 + 2.38066 N^4 $\\
0,6   &   $317.910 + 228.597 N + 59.1341 N^2 + 7.38414 N^3 + 0.318597 N^4 + 0.00727653 N^5 $\\
\hline\hline
\end{tabular}
}

\TABLE[!ht] { 
\footnotesize
\caption{MZM scheme: coefficients $b^{u}_{ij}$, cf. Eq.~(\ref{bung}).  }
\label{betaug}
\begin{tabular}{cl}
\hline\hline
\multicolumn{1}{c}{$i,j$}&
\multicolumn{1}{c}{$(N^2+N+8)^{i+j-1}\,b^{u}_{ij}$}\\
\hline \hline 
$3$,$0$  &  $-28.1481 - 6.07407\,N - 6.07407\,N^2$ \\ 
$2$,$1$  &  $-29.6296 - 29.6296\,N$ \\ 
$1$,$2$  &  $-30.8148 - 5.11111\,N - 1.7037\,N^2$ \\ 
$0$,$3$  &  $-8 - 4 N$ \\ 
\hline
$4$,$0$  &  $199.64 + 54.9404\,N + 56.2893\,N^2 + 2.69789\,N^3 + 1.34894\,N^4$ \\ 
$3$,$1$  &  $329.228 + 342.521\,N + 26.5875\,N^2 + 13.2938\,N^3$ \\ 
$2$,$2$  &  $476.927 + 191.216\,N + 81.4672\,N^2 + 1.24378\,N^3 + 0.310945\,N^4$ \\ 
$1$,$3$  &  $254.359 + 165.976\,N + 16.1265\,N^2 + 3.27131\,N^3$ \\ 
$0$,$4$  &  $68.2248 + 29.746\,N + 5.95053\,N^2$ \\ 
\hline
$5$,$0$  &  $-1832.21 - 602.521\,N - 638.341\,N^2 - 71.4848\,N^3 - 35.3533\,N^4 + 0.466938\,N^5 $ \\
         &  $+ 0.155646\,N^6$ \\ 
$4$,$1$  &  $-4128.34 - 4492.49\,N - 725.59\,N^2 - 356.03\,N^3 + 8.11729\,N^4 + 2.70576\,N^5$ \\ 
$3$,$2$  &  $-7702.2 - 4371.43\,N - 1991.42\,N^2 - 31.3703\,N^3 - 3.61396\,N^4$ \\ 
$2$,$3$  &  $-6312.85 - 4744.76\,N - 712.149\,N^2 - 143.849\,N^3 - 2.7972\,N^4 - 0.430339\,N^5$ \\ 
$1$,$4$  &  $-3175.43 - 1760.63\,N - 458.141\,N^2 - 26.2356\,N^3 - 2.65955\,N^4$ \\ 
$0$,$5$  &  $-588.999 - 356.825\,N - 57.9832\,N^2 - 4.8188\,N^3$ \\ 
\hline
$6$,$0$  &  $20770.2 + 7819.56\,N + 8488.12\,N^2 + 1340.35\,N^3 + 678.319\,N^4 + 9.91857\,N^5 $ \\ 
         &  $+3.54529\,N^6 + 0.204945\,N^7 + 0.0512362\,N^8$ \\ 
$5$,$1$  &  $59871.8 + 67528.7\,N + 15357.4\,N^2 + 7788.99\,N^3 + 135.84\,N^4 + 51.1183\,N^5 $ \\
         &  $+5.00417\,N^6 + 1.25104\,N^7$ \\ 
$4$,$2$  &  $136493. + 94965.3\,N + 46606.9\,N^2 + 2966.99\,N^3 + 844.792\,N^4 + 37.7957\,N^5 $ \\
         &  $+ 9.60758\,N^6$ \\ 
$3$,$3$  &  $151865. + 128949.\,N + 29147.1\,N^2 + 7117.85\,N^3 + 261.112\,N^4 + 54.5075\,N^5$ \\ 
$2$,$4$  &  $111193. + 76504.5\,N + 24548.1\,N^2 + 2500.48\,N^3 + 355.428\,N^4 + 2.33913\,N^5 $\\
         &  $+ 0.068185\,N^6$ \\ 
$1$,$5$  &  $41631.6 + 30068.3\,N + 7403.69\,N^2 + 1147.35\,N^3 + 43.3788\,N^4 + 2.1912\,N^5$ \\ 
$0$,$6$  &  $6838.45 + 4597.45\,N + 1180.76\,N^2 + 105.622\,N^3 + 5.48048\,N^4$ \\ 
\hline
$7$,$0$  &  $-271300. - 114181.\,N - 126851.\,N^2 - 25604.3\,N^3 - 13465.7\,N^4 - 793.197\,N^5 $ \\
& $- 259.288\,N^6 + 4.52143\,N^7 + 1.30604\,N^8 + 0.117121\,N^9 + 0.0234242\,N^{10}$ \\ 
$6$,$1$  &  $-956030. - 1.1157\,{10}^6\,N - 323182.\,N^2 - 171156.\,N^3 - 11419.1\,N^4 - 3707.12\,N^5 $ \\
& $+ 89.4175\,N^6 + 27.8058\,N^7 + 3.63425\,N^8 + 0.72685\,N^9$ \\ 
$5$,$2$  &  $-2.57728\,{10}^6 - 2.06957\,{10}^6\,N - 1.09407\,{10}^6\,N^2 - 129858.\,N^3 - 38311.9\,N^4 + 107.385\,N^5 $ \\
& $+ 97.7575\,N^6 + 29.1378\,N^7 + 6.06393\,N^8$ \\ 
$4$,$3$  &  $-3.6309\,{10}^6 - 3.40291\,{10}^6\,N - 1.00539\,{10}^6\,N^2 - 271968.\,N^3 - 9603.86\,N^4 - 1333.52\,N^5 $ \\
& $+ 103.325\,N^6 + 23.2754\,N^7$ \\ 
$3$,$4$  &  $-3.48896\,{10}^6 - 2.80102\,{10}^6\,N - 1.03092\,{10}^6\,N^2 - 132759.\,N^3 - 19077.3\,N^4 - 122.197\,N^5 $ \\
& $+ 17.666\,N^6 - 5.06322\,{10}^{-7}\,N^7 - 4.70902\,{10}^{-8}\,N^8$ \\ 
$2$,$5$  &  $-1.96674\,{10}^6 - 1.61821\,{10}^6\,N - 481214.\,N^2 - 88493.\,N^3 - 5662.61\,N^4 - 448.949\,N^5 $ \\
& $- 8.28982\,N^6 - 0.500033\,N^7$ \\ 
$1$,$6$  &  $-641804. - 491982.\,N - 154530.\,N^2 - 21660.9\,N^3 - 2089.08\,N^4 - 70.0898\,N^5 $ \\
& $- 3.10875\,N^6$ \\ 
$0$,$7$  &  $-89165.7 - 68401.1\,N - 19266.8\,N^2 - 2704.06\,N^3 - 149.873\,N^4 - 5.22722\,N^5$ \\ 
\hline \hline 
\end{tabular}
}

\TABLE[ht] { 
\footnotesize
\caption{MZM scheme: coefficients $b^{v}_{ij}$, cf. Eq.~(\ref{bvng}).  }
\label{betavg}
\begin{tabular}{cl}
\hline\hline
\multicolumn{1}{c}{$i,j$}&
\multicolumn{1}{c}{$(N^2+N+8)^{i+j-1}\;b^{v}_{ij}$}
 \\ 
\hline \hline 
$3$,$0$  &  $0$ \\ 
$2$,$1$  &  $-54.8148 - 3.40741\,N - 3.40741\,N^2$ \\ 
$1$,$2$  &  $-53.6296 - 29.6296\,N$ \\ 
$0$,$3$  &  $-20.1481 - 7.11111\,N - 1.03704\,N^2$ \\ 
\hline
$4$,$0$  &  $0$ \\ 
$3$,$1$  &  $469.334 + 41.8539\,N + 40.6028\,N^2 - 2.50221\,N^3 - 1.25111\,N^4$ \\ 
$2$,$2$  &  $720.915 + 426.443\,N - 4.51405\,N^2 - 4.50686\,N^3$ \\ 
$1$,$3$  &  $544.203 + 235.122\,N + 40.6104\,N^2$ \\ 
$0$,$4$  &  $131.416 + 69.5514\,N + 9.46842\,N^2 + 0.560179\,N^3$ \\ 
\hline
$5$,$0$  &  $0$ \\ 
$4$,$1$  &  $-5032.69 - 584.288\,N - 584.021\,N^2 - 0.0404379\,N^3 - 1.45685\,N^4 - 1.72396\,N^5 $\\
         &  $- 0.574653\,N^6$ \\ 
$3$,$2$  &  $-10619.9 - 6724.27\,N - 257.525\,N^2 - 80.9869\,N^3 - 7.06796\,N^4 - 3.10306\,N^5$ \\ 
$2$,$3$  &  $-12009.2 - 6419.67\,N - 1446.44\,N^2 + 3.57388\,N^3 - 4.26467\,N^4$ \\ 
$1$,$4$  &  $-5985.6 - 3722.89\,N - 685.886\,N^2 - 78.9515\,N^3$ \\ 
$0$,$5$  &  $-1243.21 - 746.976\,N - 168.232\,N^2 - 11.7457\,N^3 - 0.496927\,N^4$ \\ 
\hline
$6$,$0$  &  $0$ \\ 
$5$,$1$  &  $64749.3 + 9324.6\,N + 9463.87\,N^2 + 274.9\,N^3 + 128.057\,N^4 - 12.1619\,N^5 $\\
         &  $- 5.53845\,N^6 - 1.27242\,N^7 - 0.318104\,N^8$ \\ 
$4$,$2$  &  $175059. + 117375.\,N + 9362.48\,N^2 + 3195.12\,N^3 - 187.625\,N^4 - 64.2153\,N^5 $\\
         &  $- 8.29114\,N^6 - 2.37504\,N^7$ \\ 
$3$,$3$  &  $263538. + 160906.\,N + 41166.9\,N^2 - 565.238\,N^3 - 212.027\,N^4 - 16.8356\,N^5 $\\
         &  $- 6.53563\,N^6$ \\ 
$2$,$4$  &  $200359. + 139453.\,N + 28750.8\,N^2 + 3099.44\,N^3 - 96.2318\,N^4 - 16.3984\,N^5$ \\ 
$1$,$5$  &  $82989.5 + 55457.5\,N + 14428.7\,N^2 + 1410.08\,N^3 + 79.2383\,N^4$ \\ 
$0$,$6$  &  $13931.7 + 9743.27\,N + 2410.02\,N^2 + 284.777\,N^3 + 12.4061\,N^4 + 0.284188\,N^5$ \\ 
\hline
$7$,$0$  &  $0$ \\ 
$6$,$1$  &  $-943070. - 161581.\,N - 166595.\,N^2 - 9991.42\,N^3 - 4910.2\,N^4 + 92.6388\,N^5 $ \\
& $+ 13.6487\,N^6 - 15.9661\,N^7 - 5.48745\,N^8 - 0.997279\,N^9 - 0.199456\,N^{10}$ \\ 
$5$,$2$  &  $-3.12002\,{10}^6 - 2.20012\,{10}^6\,N - 271785.\,N^2 - 102493.\,N^3 + 1777.97\,N^4 + 334.971\,N^5 $ \\
& $- 192.691\,N^6 - 60.1882\,N^7 - 8.77424\,N^8 - 1.90222\,N^9$ \\ 
$4$,$3$  &  $-5.8646\,{10}^6 - 3.97094\,{10}^6\,N - 1.15071\,{10}^6\,N^2 - 22722.2\,N^3 - 3636.84\,N^4 - 814.623\,N^5 $ \\
& $- 258.054\,N^6 - 30.149\,N^7 - 7.38428\,N^8$ \\ 
$3$,$4$  &  $-6.00654\,{10}^6 - 4.59789\,{10}^6\,N - 1.11616\,{10}^6\,N^2 - 144174.\,N^3 + 2260.85\,N^4 - 257.971\,N^5 $ \\
& $- 81.4206\,N^6 - 17.4821\,N^7$ \\ 
$2$,$5$  &  $-3.73056\,{10}^6 - 2.79137\,{10}^6\,N - 817049.\,N^2 - 94705.5\,N^3 - 6992.55\,N^4 + 53.4018\,N^5 $ \\
& $- 15.7555\,N^6$ \\ 
$1$,$6$  &  $-1.2573\,{10}^6 - 969506.\,N - 282636.\,N^2 - 42677.5\,N^3 - 2754.02\,N^4 - 131.786\,N^5$ \\ 
$0$,$7$  &  $-182134. - 141435.\,N - 42673.8\,N^2 - 6033.4\,N^3 - 464.717\,N^4 - 14.301\,N^5 $ \\
& $- 0.318786\,N^6$ \\ 
\hline
\hline 
\end{tabular}
}

\TABLE[ht]{
\footnotesize
\caption{The coefficients $c^{u}_{ijk}$, $c^{v}_{ijk}$, and $c^{w}_{ijk}$, 
cf. Eqs.~(\ref{bbu}), (\ref{bbv}) and (\ref{bbw}).
The coefficients corresponding to values of $i$, $j$, and $k$ that are not 
reported are zero. 
}
\label{tabc}
\begin{tabular}{cccccc}
\hline\hline
\multicolumn{1}{c}{$i,j,k$}&
\multicolumn{1}{c}{$c^{u}_{ijk}$}&
\multicolumn{1}{c}{$c^{v}_{ijk}$}&
\multicolumn{1}{c}{$\qquad\qquad$}&
\multicolumn{1}{c}{$i,j,k$}&
\multicolumn{1}{c}{$c^{w}_{ijk}$}\\
\hline \hline
4,0,0   &   95.0752  &   0            &   &   &   \\
3,1,0   &   149.265  &   51.9181  &   & 3,0,1   &  51.9181   \\
2,2,0   &   128.028  &   123.538  &   & 2,1,1   &  $-$9.26743   \\
2,0,2   &   177.083  &$-$83.8774  &   &   &    \\
1,3,0   &   64.7434  &   118.463  &   & 1,2,1   &  $-$20.2218   \\
1,1,2   &   33.8245  &$-$33.5433  &   & 1,0,3   &   33.7500  \\
0,4,0   &   14.6233  &   31.5474  &   & 0,3,1   &   $-$2.37189  \\
0,2,2   &$-$34.2374  &   19.3311  &   & 0,1,3   &   30.3750   \\
0,0,4   &$-$16.3125  &   9.56250  &   &   &     \\
\hline
5,0,0   & $-$1002.80  &    0           &&   &      \\
4,1,0   & $-$1985.03  &  $-$646.929    &&   4,0,1   & $-$646.929  \\
3,2,0   & $-$2163.00  &  $-$1771.43    &&   3,1,1   & $-$272.688 \\
3,0,2   & $-$2308.22  &  946.573       &&   &      \\
2,3,0   & $-$1552.04  &  $-$2192.02    &&   2,2,1   & 104.725  \\
2,1,2   & $-$1003.60  &  608.979       &&   2,0,3   & $-$599.238  \\
1,4,0   & $-$671.507  &  $-$1275.70    &&   1,3,1   & 67.695  \\
1,2,2   &  290.702    &  $-$24.5584    &&   1,1,3   & $-$413.455  \\
1,0,4   & $-$166.563  &   59.9734      &&   &    \\
0,5,0   & $-$121.524  &  $-$266.862    &&   0,4,1   &   8.59444  \\
0,3,2   &  109.555    &  $-$88.119     &&   0,2,3   &   $-$49.3931 \\
0,1,4   & $-$171.978  &   46.1564      &&   0,0,5   &  $-$2.01405  \\
\hline
6,0,0   & 13083.1  &    0         &&   &    \\
5,1,0   & 31798.2  &  8542.88     && 5,0,1 &  8542.88   \\
4,2,0   & 42362.6  &  27704.8     &&   4,1,1   &  6832.08     \\
4,0,2   & 36182.0  &  $-$13182.8  &&   &    \\
3,3,0   & 38102.9  &  43067.7     &&   3,2,1   &  1007.07    \\
3,1,2   & 26890.4  &  $-$10903.4  &&   3,0,3   &  12078.4   \\
2,4,0   & 22961.9  &  36681.7     &&   2,3,1   & $-$686.727      \\
2,2,2   & 1982.93  &  $-$1712.04  &&   2,1,3   &  9048.10    \\
2,0,4   & 9154.83  &  $-$3642.81  &   &   &   \\
1,5,0   & 8272.27  & 16001.6      &&   1,4,1   &  $-$379.749     \\
1,3,2   & $-$2399.55 & 1194.47    &&   1,2,3   &  1213.86    \\
1,1,4   & 6382.20  & $-$2878.33   &&   1,0,5   &  91.8928   \\
0,6,0   & 1280.65  & 2740.04      &&   0,5,1   & $-$52.7711     \\
0,4,2   & $-$531.082  & 512.566   &&   0,3,3   &   $-$178.695  \\
0,2,4   & 777.870  &  $-$487.798 &&   0,1,5   &  106.251  \\
0,0,6   & $-$156.528 & 9.53427  &   &   &      \\
\hline\hline
\end{tabular}
}

\TABLE[ht]{
\footnotesize
\caption{
The coefficients $e^\phi_{ijk}$ and $e^{t}_{ijk}$, cf. Eq.~(\ref{etaphit}),
of the RG functions $\eta_\phi$ and $\eta_t$ in
the $\overline{\rm MS}$ and MZM schemes,
to five and six loops respectively.
The coefficients corresponding to
values of $i$, $j$, and $k$ that are not reported are zero.  }
\label{tabeta}
\renewcommand\arraystretch{0.7}
\begin{tabular}{cccll}
\hline\hline
\multicolumn{1}{c}{}&
\multicolumn{2}{c}{$\overline{\rm MS}$}&
\multicolumn{2}{c}{MZM}\\
\multicolumn{1}{c}{$i,j,k$}&
\multicolumn{1}{c}{$e^{\phi}_{ijk}$}&
\multicolumn{1}{c}{$e^{t}_{ijk}$}&
\multicolumn{1}{c}{$e^{\phi}_{ijk}$}&
\multicolumn{1}{c}{$e^{t}_{ijk}$}\\
\hline \hline
1,0,0   & 0 & $-$11/2 & 0 & $-$11/14 \\
0,1,0   & 0 & $-$5/2  & 0 & $-$5/14   \\
\hline
2,0,0   & 11/16 & 33/8 & 0.00831444  &  0.0561224 \\
1,1,0   & 5/8  & 7/2 & 0.00755858  &  0.0510204 \\
0,2,0   & 1/2  & 3   & 0.00604686  &  0.0408163 \\
0,0,2   & 3/4  & 9/2 & 0.00907029  &  0.0612245  \\
\hline
3,0,0   & $-$77/64  & $-$4675/128 & 0.000692663 & $-$0.0414663  \\
2,1,0   & $-$105/64 & $-$6375/128 & 0.000944541 & $-$0.0565449  \\
1,2,0   & $-$147/128& $-$2265/64  & 0.000661179 & $-$0.0421311  \\
1,0,2   & $-$27/32  & $-$855/32   & 0.000485764 & $-$0.0344224  \\
0,3,0   & $-$119/256& $-$893/64   & 0.00026762  & $-$0.0152562  \\
0,1,2   & 27/64     & 333/32    &  $-$0.000242882 & 0.00293313 \\
\hline
4,0,0   & 0.805664 & 273.792  & 0.00053933  & 0.0106715  \\
3,1,0   & 1.46484  & 497.803  & 0.0009806   & 0.0194027  \\
2,2,0   & 7.20703  & 521.554  & 0.00119061  & 0.027517   \\
2,0,2   & 13.0078  & 381.632  & 0.00109376  & 0.0299365   \\
1,3,0   & 8.05664  & 331.257  & 0.00085394  & 0.0217858    \\
1,1,2   & 1.75781  & 47.2145 & 0.000150938  & 0.00437099 \\
0,4,0   & 2.06909  & 87.3563 & 0.000228253  & 0.00541476 \\
0,2,2   &$-$4.13086& $-$58.0018 & $-$0.000209641  & $-$0.00713995 \\
0,0,4   &$-$3.69141& $-$14.2917 & $-$0.000116359  & $-$0.00539322 \\
\hline
5,0,0   &$-$39.5573 & $-$2842.37 & $-$0.0000621953  & $-$0.0108866 \\
4,1,0   &$-$89.9030 & $-$6459.93 & $-$0.000141353  & $-$0.0247422 \\
3,2,0   &$-$129.415 & $-$8195.78 & $-$0.000179926  & $-$0.0319839 \\
3,0,2   &$-$99.9105 & $-$4867.47 & $-$0.000107743   & $-$0.0198859 \\
2,3,0   &$-$126.293 & $-$6871.52 & $-$0.000181477  & $-$0.0276014 \\
2,1,2   &$-$21.1719 & $-$1702.98 & 0.000105679    & $-$0.00595702 \\
1,4,0   &$-$69.4979 & $-$3410.39 & $-$0.000117862  & $-$0.0137125  \\
1,2,2   &26.8948    & 622.145   & 0.000126014    & 0.00503919 \\
1,0,4   &$-$9.35655 & $-$728.248 & 0.00000742886  & 0.000156979 \\
0,5,0   &$-$13.9928 & $-$698.741 & $-$0.0000218022  & $-$0.00263618 \\
0,3,2   &2.66006    & 138.505   & $-$0.00000688957 & 0.000910744 \\
0,1,4   &$-$15.6434 & $-$540.225 & $-$0.0000318276 & $-$0.000466735 \\
\hline
6,0,0   &  & & 0.0000735165 & 0.00449246\\
5,1,0   &  & & 0.0002005    &  0.0122522\\
4,2,0   &  & & 0.000404726 &  0.0224094\\
4,0,2   &  & & 0.000370618  &  0.0177812\\
3,3,0   &  & & 0.000542556 &  0.0282151\\
3,1,2   &  & & 0.000278519  &  0.0115894\\
2,4,0   &  & & 0.000431888 &  0.0219744\\
2,2,2   &  & & 0.0000308945 &  $-$0.00073584\\
2,0,4   &  & & 0.0000838964 &  0.00391376\\
1,5,0   &  & & 0.000182469  &  0.00914719\\
1,3,2   &  & & $-$0.00000319876 & $-$0.000933258 \\
1,1,4   &  & & 0.000073068  &  0.00345242\\
0,6,0   &  & & 0.0000308055 &  0.00149629\\
0,4,2   &  & & 0.0000000587 &  0.00000318554\\
0,2,4   &  & & 0.000020701  &  0.000740282\\
0,0,6   &  & & $-$0.00000211424 &  $-$0.000129075 \\
\hline\hline
\end{tabular}
}

\TABLE[ht]{
\footnotesize
\caption{The coefficients $d^{u}_{ijk}$, $d^{v}_{ijk}$, and $d^{w}_{ijk}$, 
cf. Eqs.~(\ref{bbug}), (\ref{bbvg}) and (\ref{bbwg}).
The coefficients corresponding to values of $i$, $j$, and $k$ that are not 
reported are zero. 
}
\label{tabcg}
\renewcommand\arraystretch{0.7}
\begin{tabular}{cccccc}
\hline\hline
\multicolumn{1}{c}{$i,j,k$}&
\multicolumn{1}{c}{$d^{u}_{ijk}$}&
\multicolumn{1}{c}{$d^{v}_{ijk}$}&
\multicolumn{1}{c}{$\qquad\qquad$}&
\multicolumn{1}{c}{$i,j,k$}&
\multicolumn{1}{c}{$d^{w}_{ijk}$}\\
\hline \hline
3,0,0   & $-$0.190854   &   0            &   &   &   \\
2,1,0   & $-$0.188964   & $-$0.156841  &   & 2,0,1   & $-$0.156841  \\
1,2,0   & $-$0.100151   & $-$0.219577  &   & 1,1,1   & $-$0.0256992 \\
1,0,2   & $-$0.226757   &  0.122449 &   &   &    \\
0,3,0   & $-$0.0306122  & $-$0.0831444  &   & 0,2,1   & 0.0222978 \\
0,1,2   & $-$0.0408163  & 0.0589569  &  & 0,0,3   &  0.0181406 \\
\hline
4,0,0   &0.0837293    &   0           &   &   &   \\
3,1,0   & 0.135546    &  0.0367151    &   & 3,0,1   & 0.0367151  \\
2,2,0   & 0.123199    &  0.0941154    &   & 2,1,1   & $-$0.0168042 \\
2,0,2   & 0.162130    &  $-$0.0700545 &   &   &    \\
1,3,0   & 0.0631219   &  0.0972328    &   & 1,2,1   & $-$0.0188393 \\
1,1,2   & 0.0368576   &  $-$0.0234689 &   & 1,0,3   &  0.0339000 \\
0,4,0   & 0.0128652   &  0.0271942    &   & 0,3,1   & $-$0.000606377 \\
0,2,2   & $-$0.0293008&  0.0150557    &   & 0,1,3   &  0.0245772  \\
0,0,4   & $-$0.0191787&  0.0106478    &   &   &     \\
\hline
5,0,0   & $-$0.043871   &    0           &&   &      \\
4,1,0   & $-$0.0840225  & $-$0.0345052  &&   4,0,1   & $-$0.0345052 \\
3,2,0   & $-$0.0975887  & $-$0.0842869  &&   3,1,1   & $-$0.030022 \\
3,0,2   & $-$0.110406   & 0.0342725  &&   &      \\
2,3,0   & $-$0.0765458  & $-$0.100371  &&   2,2,1   &  $-$0.0154977 \\
2,1,2   & $-$0.0421916  & 0.0104173  &&   2,0,3   &   $-$0.0337405 \\
1,4,0   & $-$0.0323891  & $-$0.0600404  &&   1,3,1   &  $-$0.00446734 \\
1,2,2   & 0.0254609     & $-$0.0103217  &&   1,1,3   &  $-$0.0196257  \\
1,0,4   & 0.00222203    & 0.0047634  &&   &    \\
0,5,0   & $-$0.00529147 & $-$0.0126929  &&   0,4,1   &0.000106648  \\
0,3,2   & 0.00792392    & $-$0.00430159 &&   0,2,3   &0.000711959   \\
0,1,4   & $-$0.00144707 & 0.00141906    &&   0,0,5   & 0.00225136 \\
\hline
6,0,0   & 0.0278137    &    0         &&   &    \\
5,1,0   & 0.0698599    & 0.0131904   && 5,0,1 & 0.0131904   \\
4,2,0   & 0.101466     & 0.0471915  &&   4,1,1   & 0.0038472     \\
4,0,2   & 0.0893755    & $-$0.0273754  &&   &    \\
3,3,0   & 0.0994876    & 0.0831689  &&   3,2,1   & $-$0.0043311     \\
3,1,2   & 0.0600684    & $-$0.0164027  &&   3,0,3   &  0.0318928   \\
2,4,0   & 0.0618044    & 0.0799006  &&   2,3,1   &    $-$0.00342079    \\
2,2,2   & $-$0.00398625& $-$0.000641182   &&   2,1,3   & 0.0166793   \\
2,0,4   & 0.0180563    & $-$0.0120957  &   &   &   \\
1,5,0   & 0.0213326    & 0.0375477  &&   1,4,1   &   $-$0.00196111    \\
1,3,2   & $-$0.00671272& 0.000668951  &&   1,2,3   &   $-$0.00591449   \\
1,1,4   & 0.0168769    & $-$0.0124137  &&   1,0,5   &  $-$0.00290304   \\
0,6,0   & 0.0030379    & 0.00657498  &&   0,5,1   &    $-$0.000359204 \\
0,4,2   & $-$0.000589662  & 0.000193903  &&   0,3,3   & $-$0.00286458   \\
0,2,4   & 0.00331464  & $-$0.00169973  &&   0,1,5   & $-$0.00126576  \\
0,0,6   & $-$0.000644392  & 0.000712328  &   &   &      \\
\hline
7,0,0   & $-$0.0197200  &   0                 &   &   &      \\
6,1,0   & $-$0.0560554  & $-$0.0147179  & &  6,0,1   &  $-$0.0147179    \\
5,2,0   & $-$0.0935780  & $-$0.051629  &&   5,1,1   &   $-$0.0226979\\
5,0,2   & $-$0.0742718  & 0.0182722  &   &   &     \\
4,3,0   & $-$0.111549   & $-$0.0942413  &&   4,2,1   &  $-$0.0211958 \\
4,1,2   & $-$0.068364   & 0.00963234  &&   4,0,3   & $-$0.0310849  \\
3,4,0   & $-$0.0925957  & $-$0.107471  &&   3,3,1   &  $-$0.0132086 \\
3,2,2   & $-$0.0118213  & $-$0.00275039  &&   3,1,3   & $-$0.0208751  \\
3,0,4   & $-$0.0354476  & 0.016807  &     &   &   \\
2,5,0   & $-$0.0492935  & $-$0.0743529   &&   2,4,1   &  $-$0.00403603 \\
2,3,2   &  0.00447446   & $-$0.00164415  &&   2,2,3   &  0.00182088 \\
2,1,4   & $-$0.0300979  & 0.0150377  &&   2,0,5   & $-$0.00171458  \\
1,6,0   & $-$0.0147084  & $-$0.0274519  &&   1,5,1    &  $-$0.000383191\\
1,4,2   &   0.00226375  & $-$0.000784818   &&   1,3,3   & 0.00216377   \\
1,2,4   & $-$0.00407885 & 0.00239567  &&   1,1,5   &  $-$0.00290307 \\
1,0,6   &   0.00262247  & $-$0.00148025  &    &   &       \\
0,7,0   & $-$0.00184237 & $-$0.00405011  &&   0,6,1   &  0.0000382403 \\
0,5,2   &  0.000436103  & $-$0.000306989  &&   0,4,3   & $-$0.0000315805  \\
0,3,4   &  0.000962049  & $-$0.000709295  &&   0,2,5   & $-$0.000830064  \\
0,1,6   &  0.00118291   & $-$0.00098856  &&   0,0,7   &  0.0000197908 \\      
\hline\hline
\end{tabular}
}

\newpage

\begin{thebibliography}{99}


\bibitem{Wilczek-00}
F. Wilczek, {\em QCD in extreme conditions}, \hepph{0003183}.


\bibitem{Karsch-02}
F. Karsch, 
{\em Lectures on Quark Matter}, Proceedings of the 40th 
Internationale Uni\-ver\-sit\"atswochen f\"ur Kern- und
Teilchenphysik, Lecture Notes in Physics 583,
edited by W. Plessas and L. Mathelisch 
(Springer, Berlin, 2002) p. 209 
[\heplat{0106019}].

\bibitem{Rajagopal-95}
K. Rajagopal, {\em The chiral phase transition in QCD: critical phenomena and 
long wavelength pion oscillations}, in 
{\em Quark-Gluon Plasma 2}, R. Hwa ed.,
World Scientific, Singapore 1995
[\hepph{9504310}].

\bibitem{KL-03}
F. Karsch and E. Laermann, 
{\em Thermodynamics and in-medium hadron properties from lattice QCD},
to appear in {\em Quark-Gluon Plasma 3}, R. Hwa ed.,
World Scientific, Singapore
[\heplat{0305025}].

\bibitem{Peskin-80}
M.E. Peskin, 
{\em The alignment of the vacuum in theories of technicolor},
\npb{175}{1980}{197}.

\bibitem{VW-84}
C. Vafa and E. Witten, 
{\em Restrictions on symmetry breaking in vector-like gauge theories},
\npb{234}{1984}{173}.

\bibitem{LS-92}
H. Leutwyler and A. Smilga, 
{\em Spectrum of Dirac operator and role of winding number in QCD},
\prd{46}{1992}{5607}.

\bibitem{SV-95}
A. Smilga and J.J.M. Verbaarschot,
{\em Spectral sum rules and finite volume partition function 
in gauge theories with real and pseudoreal fermions},
\prd{51}{1995}{829}.

\bibitem{KSTVZ-00}
J.B. Kogut, M.A. Stephanov, D. Toublan,
J.J.M. Verbaarschot, and A. Zhitnitsky,
{\em QCD-like theories at finite baryon density},
\npb{582}{2000}{477}
[\hepph{0001171}].

\bibitem{KPSW-85}
J.B. Kogut, J. Polonyi, D.K. Sinclair, and H.W. Wyld, 
{\em Hierarchal Mass Scales in Lattice Gauge Theories with 
Dynamical, Light Fermion},
\prl{54}{1985}{1980}.

\bibitem{Kogut-87}
J.B. Kogut, {\em Simulating simple supersymmetric field theories},
\plb{187}{1987}{347}.

\bibitem{KL-99}
F. Karsch and M. L\"utgemeier,
{\em Deconfinement and chiral symmetry restoration in an SU(3) gauge theory
with adjoint fermions},
\npb{550}{1999}{449} [\heplat{9812023}].

\bibitem{ST-04}
F. Sannino and K. Tuominen,
{\em Tetracritical behavior in strongly interacting theories},
\prd{70}{2004}{034019} [\hepph{0403175}].

\bibitem{PW-84}
R.D. Pisarski and F. Wilczek, 
{\em Remarks on the chiral phase transition in chromodynamics},
\prd{29}{1984}{338}.

\bibitem{BPV-03}
A. Butti, A. Pelissetto, and E. Vicari,
{\em On the nature of the finite-temperature transition in QCD},
\jhep{08}{2003}{029} [\hepph{0307036}].

\bibitem{SD-89}
R.~Schloms and V.~Dohm,
{\em Minimal renormalization without $\epsilon$-expansion: 
Critical behavior in three dimensions},
\npb{328}{1989}{328};
{\em Minimal renormalization without $\epsilon$ expansion: 
Critical behavior above and below $T_c$},
\prb{42}{1990}{6142}.

\bibitem{Parisi-80}
G. Parisi, 
{\em Field-theoretic approach to second-order phase transitions in 
two- and three-dimensional systems},
Carg\`{e}se Lectures (1973),
\newjournal{J.\ Stat.\ Phys.\ }{JSTPB}{23}{1980}{49}.

\bibitem{tHV-72}
G. 't Hooft and M.J.G. Veltman, 
{\em Regularization and renormalization of gauge fields},
\npb{44}{1972}{189}.

\bibitem{WF-72} 
K.G.~Wilson and M.E.~Fisher,
{\em Critical exponents in 3.99 dimensions},
\prl{28}{1972}{240}.

\bibitem{LZ-77} 
J.C.~Le Guillou and J.~Zinn-Justin,
{\em Critical exponents from field theory},
\prb{21}{1980}{3976}.

\bibitem{ZJbook}
J. Zinn-Justin,
{\em Quantum Field Theory and Critical Phenomena},
fourth edition, Clarendon Press, Oxford 2001.

\bibitem{CP-PACS-01}
A. Ali Khan {\em et al}. (CP-PACS Collaboration), 
{\em Phase structure and critical temperature of two-flavor QCD 
with renormalization group improved gauge action and clover 
improved Wilson action},
\prd{63}{2001}{034502}  [\heplat{0008011}];
{\em Equation of state in finite-temperature QCD with two flavors of 
improved Wilson quarks},
\prd{64}{2001}{074510} [\heplat{0103028}].

\bibitem{KLP-01}
F. Karsch, E. Laermann, and A. Peikert,
{\em Quark mass and flavor dependence of the QCD phase transition},
\npb{605}{2001}{579}
[\heplat{0012023}].

\bibitem{GPY-81}
D.J. Gross, R.D. Pisarski, and L.G. Yaffe, 
{\em QCD and instantons at finite temperature},
\rmp{53}{1981}{43}.

\bibitem{anomalyMC}
C. Bernard, T. Blum, C. DeTar,
S.~Gottlieb, U.M.~Heller, J.E.~Hetrick,
K.~Rummukainen, R.~Sugar, D.~Toussaint, and M.~Wingate,
{\em Which chiral symmetry is restored in high temperature QCD?},
\prl{78}{1997}{598} [\heplat{9611031}].

J. B. Kogut, J.-F. Laga\"e, and D. K. Sinclair,
{\em Topology, fermionic zero modes, and flavor singlet correlators
in finite temperature QCD},
\prd{58}{1998}{054504} [\heplat{9801020}].

P. M. Vranas,
{\em The finite temperature QCD phase transition with domain wall
fermions},
\npps{83}{2000}{414} [\heplat{9911002}].

\bibitem{DPV-04}
L. Del Debbio, H. Panagopoulos, and E. Vicari,
{\em Topological susceptibility of $SU(N)$ gauge theories at 
finite temperature},
\jhep{09}{2004}{028} [\hepth{0407068}].

\bibitem{PV-r} 
A. Pelissetto and E. Vicari,
{\em Critical phenomena and renormalization-group theory},
\prep{368}{2002}{549} [\condmat{0012164}].

\bibitem{KS-01} 
H.~Kleinert and V.~Schulte-Frohlinde, 
{\em Critical Properties of $\phi^4$-Theories},
World Scientific, Singapore 2001.

\bibitem{NMB-77}
B.G.~Nickel, D.I.~Meiron, and G.A.~Baker, Jr.,
{\em Compilation of 2-pt and 4-pt graphs for continuum spin models},
Guelph University Report, 1977, unpublished.

\bibitem{KNSCL-91} 
H.~Kleinert, J.~Neu, V.~Schulte-Frohlinde, K.~G.~Chetyrkin,
and S.~A.~Larin,
{\em Five-loop renormalization group functions of ${O}(n)$-symmetric
$\phi^4$-theory and $\epsilon$-expansions of critical exponents up to 
$\epsilon^5$},
\plb{272}{1991}{39}, erratum \plb{319}{1993}{545} [\hepth{9503230}].

\bibitem{BNGM-77} 
G.A. Baker, Jr.,  B.G. Nickel, M.S. Green, and D.I. Meiron,
{\em Ising-model critical indices in three dimensions from the 
 Callan-Symanzik equation},
\prl{36}{1977}{1351}.

G.A. Baker, Jr., B.G. Nickel, and D.I. Meiron,
{\em Critical indices from perturbation analysis of the 
Callan-Symanzik equation},
\prb{17}{1978}{1365}.

\bibitem{AS-95} 
S.~A.~Antonenko and A.I.~Sokolov,
{\em Critical exponents for a three-dimensional $O(n)$-symmetric model with 
$n>3$},
\pre{51}{1995}{1894} [\hepth{9803264}].

\bibitem{Kawamura-88}
H.~Kawamura,
{\em Renormalization-group analysis of chiral transitions},
\prb{38}{1988}{4916}; erratum \prb{42}{1990}{2610}.

\bibitem{Kawamura-98}
H.~Kawamura, 
{\em Universality of phase transitions of frustrated antiferromagnets},
\newjournal{J.\  Phys.\ : Condens.\ Matter}{JCOME}{10}{1998}{4707}
[\condmat{9805134}].

\bibitem{PRV-01}
A.~Pelissetto, P. Rossi, and E.~Vicari, 
{\em Critical behavior of frustrated spin models with noncollinear order},
\prb{63}{2001}{140414} [\condmat{000739}].

\bibitem{CP-04}
P. Calabrese and P. Parruccini, 
{\em Five-loop $\epsilon$ expansion for $O(n)\otimes O(m)$ spin models},
\npb{679}{2004}{568} [\condmat{0308037}]. 

\bibitem{highorder}
{\em Large-Order Behaviour of Perturbation Theory},
J.C.~Le~Guillou and L. Zinn-Justin eds.,
North-Holland, Amsterdam 1990.

\bibitem{BD-84}
M.C.~Berg\`ere and F. David,
{\em Ambiguities of renormalized $\phi^4_4$ field theory and the singularities 
of its Borel transform},
\plb{135}{1984}{412}.

\bibitem{CPRV-02}
M. Campostrini, A. Pelissetto, P. Rossi, and E. Vicari,
{\em 25th-order high-temperature expansion results for 
three-dimensional Ising-like systems on the simple-cubic lattice},
\pre{65}{2002}{066127} [\condmat{0201180}].

\bibitem{DB-03}
Y. Deng and H.W.J. Bl\"ote,
{\em Simultaneous analysis of several models in the three-dimensional 
Ising universality class},
\pre{68}{2003}{036125}.

\bibitem{CHPRV-01} 
M.~Campostrini, M.~Hasenbusch, A.~Pelissetto, P.~Rossi, and E.~Vicari, 
{\em Critical behavior of the three-dimensional $XY$ universality class},
\prb{63}{2001}{214503} [\condmat{0010360}].

\bibitem{CPV-03}
P. Calabrese, A. Pelissetto, and E. Vicari,
{\em Multicritical phenomena in $O(n_1)\oplus O(n_2)$-symmetric theories},
\prb{67}{2003}{054505} [\condmat{0209580}].

\bibitem{superc}
B. I. Halperin, T. C. Lubensky, and S. K. Ma,
{\em First-order phase transitions in superconductors and 
smectic-$A$ liquid crystals},
\prl{32}{1974}{292}.

C.W. Garland and G. Nounesis, 
{\em Critical behavior at nematic--smectic-$A$ phase transitions},
\pre{49}{1994}{2964}.

H. Kleinert, {\em Gauge Fields in Condensed Matter},
World Scientific, Singapore 1989.

K. Kajantie, M. Karjalainen, M. Laine, and J. Peisa,
{\em Masses and phase structure in the Ginzburg-Landau model},
\prb{57}{1998}{3011} [\condmat{9704056}].

S. Mo, J. Hove, and A. Sudb\o,
{\em Order of the metal-to-superconductor transition},
\prb{65}{2002}{104501} [\condmat{0109260}].

\bibitem{CPPV-04}
P. Calabrese, P. Parruccini, A. Pelissetto, and E. Vicari, 
{\em Critical behavior of $O(2)\otimes O(N)$ symmetric models},
Phys. Rev. {\bf B 70} (2004) 174439 [\condmat{0405667}].

\bibitem{he3}
M. De Prato, A. Pelissetto, and E. Vicari, 
{\em Normal-to-planar superfluid transition in ${}^3$He },
Phys. Rev. {\bf B 70} (2004) ????? [\condmat{0312362}].

\bibitem{CPS-02}
P. Calabrese, P. Parruccini, and A.I. Sokolov,
{\em Chiral phase transitions: Focus driven critical behavior in 
systems with planar and vector ordering},
\prb{66}{2002}{180403(R)} [\condmat{0205046}].

\bibitem{CPV-04}
P. Calabrese, A. Pelissetto, and E. Vicari,
{\em Multicritical behavior in frustrated spin systems 
with noncollinear order}, to appear in Nucl. Phys. {\bf B} (2005)
[\condmat{0408130}].

\bibitem{CP-04-unxum}
P. Calabrese and P. Parruccini,
{\em Five-loop $\epsilon$ expansion for $U(n)\otimes U(m)$ models: 
finite-temperature phase transition in light QCD},
\jhep{05}{2004}{018} [\hepph{0403140}]

\bibitem{CPV-00}
J.M. Carmona, A. Pelissetto, and E. Vicari,
{\em $N$-component Ginzburg-Landau Hamiltonian with cubic anisotropy: 
A six-loop study},
\prb{61}{2000}{15136}.

\bibitem{PV-04}
A. Pelissetto and E. Vicari, 
{\em Interacting $N$-vector order parameters with $O(N)$ symmetry},
to appear in {\em Condens. Matt. Phys. (Ukraine)} (2005) 
[\hepth{0409214}].

\bibitem{CHPRV-02} 
M. Campostrini, M. Hasenbusch, A. Pelissetto,
P. Rossi, and E. Vicari, 
{\em Critical exponents and equation of state of the three-dimensional 
Heisenberg universality class},
\prb{65}{2002}{144520} [\condmat{0110336}].

\bibitem{softening}
Y.~Imry and M.~Wortis, 
{\em Influence of quenched impurities on first-order phase transitions},
\prb{19}{1979}{3580}.

M.~Aizenman and J.~Wehr, 
{\em Rounding of first-order phase transitions in systems 
with quenched disorder},
\prl{62}{1989}{2503}.

K.~Hui and A.~Nihat Berker, 
{\em Random-field mechanism in random-bond multicritical systems},
\prl{62}{1989}{2507}.

J.~Cardy, 
{\em Effect of random impurities on fluctuation-driven 
first-order transitions},
\jpha{29}{1996}{1897}.




\end{thebibliography}
\end{document}